%% file: CAPP_final.tex
\documentclass[12pt]{article}
\usepackage{amsmath}
\usepackage{graphicx}
\usepackage[round]{natbib} \setcitestyle{aysep={}}
\usepackage{url} 

\newcommand{\blind}{0}

\usepackage[normalem]{ulem}
\usepackage[dvipsnames]{xcolor} 
\usepackage[T1]{fontenc}
\usepackage[english]{babel}
\usepackage{graphicx}			
\usepackage{subfig}             
\usepackage{authblk}
\usepackage{amsmath} 			
\usepackage{amssymb} 			
\usepackage{calligra}			
\usepackage{calrsfs}
\usepackage{bm}					
\usepackage{bbm}
\usepackage{amsthm} 			
\usepackage{listings}			
\usepackage{multicol}
\usepackage{multirow} 
\usepackage{enumitem}			
\usepackage{newtxmath}			
\usepackage[fleqn]{mathtools}
\usepackage{tabularx,ragged2e}
\usepackage[ruled,vlined]{algorithm2e} 
\usepackage{adjustbox}
\usepackage{lineno} 
\usepackage{hyperref}
\usepackage[title]{appendix}

\hypersetup{
    colorlinks  = true,
    linkcolor   = black,
    urlcolor    = black,
    citecolor   = blue
}

\usepackage[nameinlink]{cleveref} 
\DeclareRobustCommand{\bbone}{\text{\usefont{U}{bbold}{m}{n}1}}

\newcommand{\indep}{\perp \!\!\! \perp}

\newcommand{\mytilde}[1]{\mathrel{\overset{\makebox[0pt]{\mbox{\normalfont\tiny\sffamily #1}}}{\sim}}}

\newcommand{\tilda}{~}
\newenvironment{nalign}{
	\begin{equation}
	\begin{aligned}
}{
	\end{aligned}
	\end{equation}
	\ignorespacesafterend
}

\newcommand{\GWish}[2]{\operatorname{G-Wishart}\left(#1, #2\right)}
\definecolor{codegreen}{rgb}{0,0.6,0}
\definecolor{codegray}{rgb}{0.5,0.5,0.5}
\definecolor{codepurple}{rgb}{0.58,0,0.82}
\definecolor{codeorange}{RGB}{216,101,0}
\definecolor{backcolour}{rgb}{0.95,0.95,0.92}

\lstdefinestyle{mystyle}{
	backgroundcolor=\color{backcolour},   
	commentstyle=\fontfamily{pcr}\color{codegreen},
	keywordstyle=\color{codeorange},
	numberstyle=\tiny\color{black},
	stringstyle=\color{violet},
	basicstyle=\footnotesize\fontfamily{pcr}\color{blue},
	breakatwhitespace=false,         
	breaklines=true,                 
	captionpos=b,                    
	keepspaces=true,                 
	showspaces=false,                
	showstringspaces=false,
	showtabs=false,                  
	tabsize=4
}
\lstdefinelanguage{C++} {
	morekeywords={class, public, private, explicit, inline
	},
	sensitive=false,
	morecomment=[l]{//},
	morecomment=[s]{/*}{*/},
	morestring=[b]" 
}

\renewcommand{\P}{\mathbb{P}}

\newcommand{\tr}{\mbox{tr}}

\newcommand{\T}{^{T}} 
\newcommand{\ats}{^{[s]}} 

\newcommand{\bmK}{\bm{K}}
\newcommand{\bmW}{\bm{W}}

\newcommand{\bmPhi}{\bm{\Phi}}

\newcommand{\CcB}{\mathcal{B}}

\newcommand{\CcE}{\mathcal{E}}

\newcommand{\CcG}{\mathcal{G}}

\newcommand{\bmwidetW}{\bm{\widetilde{W}}}
\newcommand{\bmwidetPhi}{\bm{\widetilde{\Phi}}}

\def\spacingset#1{\renewcommand{\baselinestretch}%
{#1}\small\normalsize} \spacingset{1}
\date{}

\begin{document}

\if0\blind
{
  \title{\bf Learning Block Structured Graphs in Gaussian Graphical Models}
  \author{Alessandro Colombi\thanks{
    The research of the third and fourth author has been partially supported by a grant from Università Cattolica del Sacro Cuore, Italy (track D1).}\hspace{.2cm}\\
    Department of Economics, Management and Statistics, Università degli studi di Milano-Bicocca, Italy\\
    and \\
    Raffaele Argiento \\
    Department of Economics, Università degli studi di Bergamo, Italy\\
    and \\
    Lucia Paci\\
    Department of Statistical Sciences, Università Cattolica del Sacro Cuore, Milan, Italy\\
    and\\
    Alessia Pini\\
    Department of Statistical Sciences, Università Cattolica del Sacro Cuore, Milan, Italy}
  \maketitle
} \fi

\if1\blind
{
  \bigskip
  \bigskip
  \bigskip
  \begin{center}
    {\LARGE\bf Learning Block Structured Graphs in Gaussian Graphical Models}
\end{center}
  \medskip
} \fi

\bigskip
\begin{abstract}
A prior distribution for the underlying graph is introduced in the framework of Gaussian graphical models. Such a prior distribution induces a block structure in the graph's adjacency matrix, allowing learning relationships between fixed groups of variables.
A novel sampling strategy named Double Reversible Jumps Markov chain Monte Carlo is developed for learning block structured graphs under the conjugate G-Wishart prior. 
The algorithm proposes moves that add or remove not just a single edge of the graph but an entire group of edges. The  method is then applied to smooth functional data. The classical smoothing procedure is improved by placing a graphical model on the basis expansion coefficients, providing an estimate of their conditional dependence structure. Since the elements of a B-Spline basis have compact support, the conditional dependence structure is reflected on well-defined portions of the domain. A known partition of the functional domain is exploited to investigate relationships among portions of the domain and improve the interpretability of the results. Supplementary materials for this article are available online.
\end{abstract}

\noindent%
{\it Keywords:} Bayesian statistics;  conditional independence; functional data analysis; G-Wishart prior; Reversible jump MCMC
\vfill

\newpage
\spacingset{1.5} 

\input{1-Introduction}

\input{2-ProposedModel}

\input{3-BlockDoubleReversibleJumpsMCMC}

\input{4-SimulationStudy}
\input{5-Applications}

\input{6-Discussion}

\input{ackno}
\bibliographystyle{jasa3}
\bibliography{biblio}

\input{Supplementary}

\end{document}

%% file: 1-Introduction.tex
\section{Introduction}
\label{section:intro}

Probabilistic graphical modeling is a powerful tool for studying the dependence structure among variables. The approach relies on the concept of conditional independence between variables, which is described through a map between a graph and a family of multivariate probability models. When the family of probability distributions is Gaussian, such models are known as Gaussian graphical models \citep{lauritzen1996}. These models have been widely applied in many research fields to infer various types of networks, including in genomics \citep{peterson.stingo.vannucci2015, castelletti2020}, health surveillance \citep{dobra2011bayesian} and finance \citep{scottj.gcarvalhoc.m2008,  wang2015scaling}. 

Let $\bm{Y}$  
be a $p$-random vector distributed as a multivariate normal distribution with zero mean and precision matrix $\bm{K}$, i.e.,  $\mbox{N}_{p}\left(\bm{0},\bm{K}^{-1}\right)$. 
Under this normality assumption, the conditional independence relationship between variables can be represented in terms of the null elements of the precision matrix $\bm{K}$. 
Specifically, let $G=(V,E)$ be an undirected graph, where $V=\{1,\dots,p\}$ is the set of $p$ nodes, and $E$ is the set of undirected edges between the nodes. Each node of the graph is associated with one of the variables of interest, and the edges describe the structure of the non-zero elements of the precision matrix. The absence of a link between two vertices is equivalent to the conditional independence of the corresponding variables, given all the others. Moreover, the entries of the precision matrix corresponding to the two variables are null. 

Usually, the structure of the underlying graph is unknown and needs to be estimated based on the available data
: this is referred to as (graph) structural learning. 
In a Bayesian framework, the specification
of a prior on the graph space 
and, conditionally on the graph, a prior on the precision
matrix is required. 
A common practice is to choose a discrete uniform distribution over the graph space $\mathcal{G}$, i.e., the space of all possible undirected graphs with $p$ nodes. This is appealing for its simplicity but  
 it is not a convenient choice to encourage sparsity, as it assigns most of the probability mass to graphs with a "medium" number of edges \citep{jonesb.carvalhoc.dobraa.hansc.carterc.westm.2005}. 
As an alternative, the prior over $\mathcal{G}$ is induced by assuming  independent  $\operatorname{Bernoulli}(\theta_e)$ priors for each edge. 
The $\operatorname{Bernoulli}$ parameters $\theta_e$ may differ from edge to edge, but usually, a common value $\theta$ is assigned.
For example, \cite{jonesb.carvalhoc.dobraa.hansc.carterc.westm.2005} suggested setting 
$\theta = 2/(p-1)$ to encourage  sparsity in the graph;  \citet{scottj.gcarvalhoc.m2008} placed instead a Beta hyperprior on $\theta$, a solution known  as the multiplicity correction prior.
Similarly, \citet{scutari2013prior} described a multivariate Bernoulli distribution where edges are not necessarily independent.

A common feature of the graph priors mentioned above is that they are non-informative since the only type of prior belief they elicit in the model is the expected sparsity rate of the graph. 
In this work, instead, we develop a prior on the graph space that aims to be informative,
according to prior information available for the application at hand.
Since the graph describes the conditional dependence structure of variables involved in complex and, usually, high-dimensional phenomena, it is unrealistic to assume that prior knowledge is available for any one-to-one relationships between the observed quantities. 
Rather, it is more reasonable to envision that variables are grouped into smaller subsets. 
This is common in biological applications where the groups may be families of bacteria \citep{osborne2021latent},
or genomics where 
groups of genes are known to be part of a common process \citep{yook}. Also in market basket analysis products and customers can  be easily grouped \citep{giudicip.castelor.2003}.

Additionally, in some applications, the variables of interest may have a natural ordering, leading to groups of nodes that are contiguous with respect to such ordering. In this case, if prior information is available about a possible partition of the nodes, the modeling assumptions must reflect that nodes cannot be re-labeled. 
 For example,  in spectrometric data analysis, the goal is usually to investigate relationships among the substances within a
compound by observing its spectrum, which can be represented as a continuous function of the wavelength. To this end, a nonparametric regression coupled with a Gaussian graphical model  on basis expansion
coefficients can be employed for smoothing the data, providing an estimate of their conditional independence structure \citep{codazzi2021gaussian}. Since the elements of a B-Spline basis have compact support, the conditional independence structure of the smoothing coefficients is reflected on portions of the spectrum, that are known a priori to be grouped in intervals of chemical interest. Here, the nodes representing the spline coefficients  are naturally ordered and the fixed groups of nodes turn out to be contiguous. Thus, a prior on the graph should elicit such relevant features.

In this paper, we propose a class of prior distributions on the graph space that leverages information on the groups of nodes and encodes a block structure in the adjacency matrix associated with  the graph.  Our approach consists of mapping the  graph $G$ into a block structured multigraph representation $G_B$, where blocks of edges are represented by a single edge between two groups of nodes. Then, independent Bernoulli priors are assumed for the edges of the multigraph $G_B$. In other words, we allow nodes in different groups to be only fully connected or not connected at all. As a result, posterior learning aims at 
discovering 
the underlying pattern between groups of nodes, based on
of the available data. We call this novel class of priors as \textit{block graph priors}.

Bayesian posterior inference on the graph is usually performed through Markov chain Monte Carlo (MCMC) under the conjugate G-Wishart  prior distribution on the precision matrix \citep{roveratoa.2002, a.ataykayish.massam2005}. However, posterior computation  is expensive for general non-decomposable graphs for two main reasons.
Firstly, note that the cardinality of the graph space is $\lvert \mathcal{G} \lvert = 2^{\binom{p}{2}}$, i.e., it is large even if a moderate number of variables $p$ is included.
In practice, it can not be explicitly enumerated but one needs to rely on search algorithms to explore it and learn which edges should be included or not.  However, even when the number of nodes is limited, it may be  difficult to identify high posterior probability regions of the graph space.

A second challenge for developing efficient methods for structural learning is due to the presence of the G-Wishart prior distribution, which is defined conditionally on a graph $G$, and it is known only up to an intractable normalizing constant. 
Explicit formulas do exist for special cases such as complete or decomposable graphs \citep{decomposable1993} which, however, are hard to justify from an applied side and increasingly restrictive as the number of nodes increases. \citet{uhler2018exact} provided a recursive expression for the normalizing constant, but the procedure is computationally efficient only for some specific types of
graphs. In practice, the normalizing constant 
is usually evaluated through numerical approximations such as the importance sampler \citep{roveratoa.2002, dellaportas2003bayesian}, the Monte Carlo approximation \citep{a.ataykayish.massam2005},  and the Laplace approximation  \citep{moghaddam2009accelerating,lenkoskia.dobraa.2011}. 
Unfortunately, these methods become unstable with an increasing number of nodes (see  \citealt{jonesb.carvalhoc.dobraa.hansc.carterc.westm.2005} and \citealt{wangh.lis.z.2012} for further details). 
Recent solutions have been proposed in the literature; for example, \cite{wangh.lis.z.2012} leverages on the partial analytical structure of the G-Wishart distribution while \citet{BDgraphpackage} rely on an approximation of the ratio of two  normalizing constants arising when  two models are compared.  Also, \citet{willemWWA}  
used a delayed acceptance MCMC \citep{delayedAcceptance} coupled with an  informed proposal  distribution \citep{zanella} on the graph space to enable embarrassingly parallel computation. %
However, all these approaches are suited for 
comparing models whose graphs differ by a single edge, and so they are inappropriate to address block structural learning. Rather, an MCMC method that modifies more than one link at a time is needed in our setting. 

In this work, we introduce a Reversible Jump MCMC sampler \citep{giudici1999decomposable, dobra2011bayesian}, defined over the joint space of the graph and the precision matrix, that leverages the structure induced by the block graph prior. In particular, we generalize the procedure of \citet{alexlenkoski2013} so that the algorithm modifies an entire block of edges at each step of the chain to guarantee a block structure of the adjacency matrix associated with the graph. Moreover, the Reversible Jump algorithm  is coupled with the Exchange algorithm \citep{murray2012mcmc}, consisting of a second reversible move, to avoid the calculation of the G-Wishart normalizing constant. We refer to the novel sampling method as the \emph{Block Double Reversible Jump} (BDRJ) algorithm.
As a result, the algorithm builds a Markov chain that visits only the subspace of block structured graphs, which is, in general, much smaller relative to the original graph space and allows us to infer the relationships among the fixed groups of nodes. 

The remainder of the paper is organized as follows.  \Cref{section:blockprior} introduces  the class of block structured graph priors and \Cref{section:BDRJ} provides the novel sampling strategy. In \Cref{section:simulation} we present a simulation study along with a comparison against existing approaches. \Cref{section:fruit_purees} illustrates our method for functional data analysis. We conclude with a brief discussion in \Cref{section:discussion}.

%% file: 2-ProposedModel.tex
\section{Block Structured Graph prior}
\label{section:blockprior}
\subsection{From a graph to a block multigraph}
\label{subsection:multigraph}
Let $\bm{Y}$  
be a $p$-random vector distributed as a multivariate normal distribution with zero mean and precision matrix $\bm{K}$, i.e.,  $\mbox{N}_{p}\left(\bm{0},\bm{K}^{-1}\right)$; without loss of generality, we assume here $\bm{Y}$ to be centered around zero.  
Let $G=(V,E)$ be an undirected graph, where $V=\{1,\dots,p\}$ is the set of $p$ nodes and $E\subset\left\{~(i,j)~\lvert~i<j,\tilda i,j\in V~\right\}$ is the set of undirected edges between the nodes. 
$\bm{Y}$ is said to be Markov with respect to $G$ if, for any edge $(i,j)$ that does not belong to $E$, the $i$-th and $j$-th variables of $\bm{Y}$ are conditionally independent given all the others, i.e., $ Y_{i}\indep Y_{j} \tilda\lvert\tilda \bm{Y}_{-(ij)} $, where $\bm{Y}_{-(ij)}$ is the random vector containing all elements in $\bm{Y}$ except the $i$-th and the $j$-th.
Under the normality assumption, the conditional independence relationship between variables has an equivalent representation in terms of the null elements of the precision matrix $\bm{K}$.
Specifically, each node is associated with one of the variables of interest, and edges describe the structure of the non-zero elements of the precision matrix. The absence of a link between two nodes means that the two corresponding variables are conditionally independent, given all the others, and the corresponding entry of the precision matrix is zero. Hence, the following equivalence provides an interpretation of the graph
\begin{equation}\nonumber
    Y_{i}\indep Y_{j} \tilda\lvert\tilda \bm{Y}_{-(ij)} \iff (i,j)\notin E \iff k_{ij} = 0,
    \label{eqn:GGM_equivalence}
\end{equation}
where $k_{ij}$ is the entry $(i,j)$ of the precision matrix $\bm{K}$.
The graph $G$ is usually unknown and must be learned from the data. In a Bayesian framework, it is considered as a random variable having values in $\mathcal{G}$, i.e., the space of all possible undirected graphs with $p$ nodes. 
The starting point for our proposed model assumes that the $p$ observed variables are grouped in $M$ mutually exclusive groups that are known a priori. Each group has cardinality $n_{i}$ and $\sum_{i = 1}^{M}n_{i} = p$. We admit the possibility of having some $n_{i} = 1$, as long as $M<p$.

In our setting, the groups are given, and the  adjacency matrix of the underlying graph has to satisfy a precise block structure. To accomplish that, we define
a new space of undirected graphs whose nodes represent the groups of variables and edges represent the  relationships between them.
Namely, let $V_{B}=\{B_{1},\dots,B_{M}\}$ be a partition of $V$ in $M$ groups that is available a priori. 
Then, we define $G_{B}=(V_{B},E_{B})$ to be an undirected graph whose nodes are the sets $B_{k}, k=1,\dots,M$ and that allows for a self-loop for node $k$ if $n_{k}>1$. Namely, the set of edges $E_B$ is given by
\[
    E_{B}\subset
    \Bigl\{
        (l,m)\tilda\lvert\tilda l,m \in V_{B},\tilda \wedge \tilda l<m, \quad
        (l,l)\tilda\lvert\tilda l\in V_{B}, \tilda \wedge \tilda n_{l}>1
    \Bigr\}.
    \label{eqn:Eb}
\]
Graphs that have self-loops are called \textit{multigraphs}. 
We denote by $\mathcal{G}_{B}$  the set of all possible multigraphs $G_{B}$ having $V_{B}$ as a set of nodes. 

\begin{figure}[h]
	\centering
	\includegraphics[width=\linewidth]{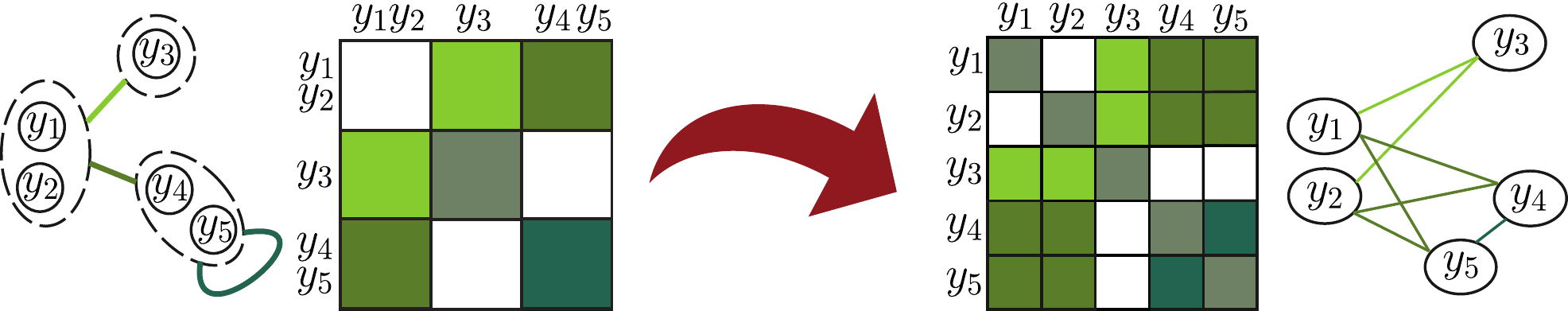}
	\caption[$\rho$ map]{The map from multigraph $G_{B}\in\mathcal{G}_{B}$ (left panel) to its block structured form $G\in\mathcal{B}$ (right panel). }
	\label{fig:mapgga}
\end{figure}

To clarify the relationship between $\mathcal{G}_B$ and $\mathcal{G}$, consider $G_{B}\in \mathcal{G}_{B}$ and $G \in \mathcal{G}$. By definition, the set of nodes of the multigraph $G_B$ is obtained by grouping  the nodes of the graph $G$. 
The following map defines a relationship between the two sets of edges. 
Let $\rho: \mathcal{G}_{B} \rightarrow \mathcal{G}$, such that  $G_{B}=(V_{B},E_{B})\mapsto G=(V,E)$ by the following transformations:
\begin{nalign}
&V = \{1,\dots,p\} = \bigcup_{k=1}^M B_k\\
&\text{if }(h,k)\in E_{B} \Rightarrow ~ (i,j)\in E ~\forall i \in B_{h}, ~\forall j \in B_{k}\\
&\text{if }(h,k)\notin E_{B} \Rightarrow ~ (i,j)\notin E ~\forall i \in B_{h}, ~\forall j \in B_{k}
\label{eqn:rho_def}
\end{nalign} 
A visual representation of this mapping is given in \Cref{fig:mapgga}. 
Once $\rho$ is set we can associate each $G_{B}$ in $\mathcal{G}_{B}$ to one and only one $G$ in $\mathcal{G}$ since $\rho$ is injective. We refer to $G_{B}$ as the multigraph form of $G$.

Nevertheless, $\rho$ is not surjective which implies that there are graphs $G$ that do not have a representative in $\mathcal{G}_{B}$. Indeed, only those graphs whose adjacency matrix has a particular block structure can be represented in a multigraph form. The non-surjective map is the key ingredient to define a subset of $\mathcal{G}$ of block structured graphs that satisfy the modeling assumptions.
Let $\mathcal{B}$ be  the image of $\rho$, i.e., the subset of $\mathcal{G}$ containing all the graphs having $p$ nodes and a block structure consistent with $V_B$. Moreover, $\rho: \mathcal{G}_{B} \rightarrow \mathcal{B}$ is a bijection, which means that every graph $G \in \mathcal{B}$ is associated to its representative $G_{B} \in \mathcal{G}_{B}$ via $\rho^{-1}$. We say that $G \in \mathcal{B}$ is the block graph representation of the multigraph $G_{B} \in \mathcal{G}_{B}$. This  representation of block graphs allows us to work in a space where we can use standard tools of graphical analysis. 

In a different setting, \citet{cremaschi2021seemingly} employ a block structure of the graph that is similar to ours. In their work, the multigraph is used to describe the conditional dependence structure across Markov processes, while the larger graph is used to capture the dependence at the observation level. However, unlike our approach, the self-loops in the multigraph  are assumed to be known and not learned from the data, due to the nature of their application. Hence, they end up with a larger graph which always  assumes the presence of $M$ cliques.

\subsection{Prior on the  graph space}
\label{subsection:prior}
The map described in Section \ref{subsection:multigraph} allows us to introduce a class of priors on $\CcG$ that encodes the knowledge about the partition of the nodes.
A customary choice in the literature is to assume independent Bernoulli priors for edges in $\CcE$, the set of all possible edges.  Such an assumption seems reasonable only if one may assume a priori independence between the edges.
Nevertheless, to induce a block structure in the adjacency matrix of the graph,  independent Bernoulli priors for the edges would not be
appropriate. 

The class of priors we propose is built on two assumptions: (i) the nodes have a natural ordering and their groups are known a priori; (ii) a zero mass probability must be placed to those graphs in $\CcG$ where nodes in the same groups are not fully connected or not connected at all, i.e., to all graphs that belong to $\CcG\backslash\CcB$.
Based on these assumptions, we need to specify the probability of graphs that are in $\CcB$, which can be represented through a multigraph $G_B\in\CcG_B$ using Equation \eqref{eqn:rho_def}. 
Each graph in $\CcG_B$ can be thought as an undirected graph having $M$ nodes and possible self-loops. Therefore, it is reasonable to assume that a prior on $\CcG_B$ can be defined by assigning independent Bernoulli priors to its edges.
Finally, the prior probability of each graph in $\CcB$ is set to be equal to the probability of its representative in $\CcG_B$, which can be obtained by applying the $\rho^{-1}$ map. 
Namely,
\begin{equation}
    \pi(G) \propto \begin{cases}
    ~\pi_B\left(\rho^{-1}(G)\right),  &\qquad \text{if  } G \in \mathcal{B},  \\
    ~0, &\qquad \text{if  }G \in \mathcal{G} \backslash\mathcal{B}, \\
    \end{cases}
    \label{eqn:truncated}			
\end{equation}
where $\pi_B\left(\rho^{-1}(G)\right) = \pi_{B}\left(G_B \right)=\theta^{\lvert E_B\lvert}(1 - \theta)^{\binom{M}{2} - \lvert E_B \lvert}$ is the $\operatorname{Bernoulli}$ prior over the set $\CcG_B$, where each link has prior probability of inclusion $\theta$, which is fixed a priori.
We refer to the prior distribution in Equation \eqref{eqn:truncated}  as \textit{block-Bernoulli prior.} In particular,  the prior is composed of two ingredients: $\pi_B(\cdot)$, that is a prior on the space $\CcG_B$ and $\rho^{-1}$ that is deterministic. This implies that the same construction is valid even if $\pi_B(\cdot)$ is replaced by any other prior distribution for graphical models. The only constraint is that it must be a probability distribution over $\CcG_B$, not over $\CcG$. We define  the resulting  class of priors to as the \textit{block graph priors}.

%% file: 3-BlockDoubleReversibleJumpsMCMC.tex
\section{Block Double Reversible Jump algorithm}
\label{section:BDRJ}

A popular choice as a prior for the precision matrix $\bmK$, conditional on the graph $G$, is
the G-Wishart distribution, introduced by \citet{roveratoa.2002} to deal with non-decomposable graphs. 
Following \citet{mohammadi.wit2015}, we work with a Shape-Inverse Scale parametrization of the G-Wishart distribution, that is, 
we say that $\bm{K}\mid G \mytilde{~} \operatorname{G-Wishart}(b,D)$ if its density is given by
\begin{equation}\nonumber
    P\left(\bm{K}~\lvert~ G \right) = I_{G}\left(b, D\right)^{-1} \lvert \bm{K}\rvert^{\frac{b - 2}{2}} \exp\left\{ - \frac{1}{2}\tr\left(\bm{K}D\right)\right\}
    \tilda\bbone_{\mathbb{P}_{G}},
    \label{eqn:GWish_def}
\end{equation}
where $\tr$ is the trace operator, $b > 2$ is the shape parameter, $D$ is the inverse scale matrix which is symmetric and positive definite, and $\mathbb{P}_{G}$ is the space of all $p\times p$ symmetric and positive definite matrices 
that are Markov with respect to $G$. Finally,
\begin{equation}\nonumber
    I_{G}(b,D) = \int_{\mathbb{P}_{G}}\lvert \bm{K}\rvert^{\frac{b - 2}{2}} \exp\left\{ - \frac{1}{2}\tr\left(\bm{K}D\right)\right\}~d\bm{K},
    \label{eqn:IGconst_def}
\end{equation}
is the normalizing constant.  
In this work, $b$ and $D$ are fixed hyperparameters. Let $\bm{y}=(\bm{y}_1, \dots, \bm{y}_n)$ be a iid sample of size $n$ from a $\mbox{N}_p\left(\bm{0}, \bm{K}^{-1}\right)$, where the precision matrix $\bm{K}$ is Markov with respect to the graph $G$. Thanks to conjugacy, the full conditional distribution of $\bm{K}$ is 
$\bmK\tilda\lvert\tilda G, \bm{y} \sim \GWish{b+n}{D+U}$, where $U=\bm{y}\top\bm{y}$.

The goal of Bayesian structural learning is to compute the posterior distribution 
\[
    P(G\mid\mathbf{y})\tilda = \tilda
        \int_{\P_G}p(\mathbf{y}\mid\bmK) p(\bmK\mid G)\pi(G)d\bmK 
    \tilda\propto\tilda
    \pi(G)\frac{I_G(d+n, D+U)}{I_G(d, D)},
    \label{eqn:apx_posteriorG}
\]
which depends on the ratio of the posterior and the prior G-Wishart normalizing constants. 

As anticipated in the Introduction, posterior  inference with the G-Wishart distribution is challenging since the joint posterior distribution of the graph and the precision matrix is doubly intractable \citep{murray2012mcmc}. Indeed, the normalizing  constant $I_{G}(b,D)$ does not have a simple analytical form for general
non-decomposable graphs, making the computation  of a Metropolis-Hastings acceptance probability not feasible. To address this issue, Monte Carlo \citep{a.ataykayish.massam2005} and Laplace   approximations \citep{moghaddam2009accelerating,lenkoskia.dobraa.2011} have been introduced. Alternatively, \citet{mohammadi2017ratio} proposed an approximation of the ratio  $I_{G}(b,D)/I_{G'}(b,D)$,  which is, in practice, the quantity required in the computation of the acceptance-rejection probability of a proposed graph $G'$. However, all the aforementioned methods are not able to exploit prior information about the block structure of the graph since they are suited to modify only one edge at each step of the MCMC algorithm. Rather, to ensure a block structure of the graph compatible with our prior beliefs, edges can not be modified at will, at least not in the space $\mathcal{B}$. To our knowledge, there are not theoretically grounded methods available in the literature to compute the G-Wishart normalizing constant ratio in our setting.  

For this reason, we move a step forward and develop a Block Reversible Jump Markov chain Monte Carlo sampler \citep{giudici1999decomposable, dobra2011bayesian} defined over the joint space of graph and precision matrix. 
It generalizes the procedure by \cite{alexlenkoski2013} in such a way that it modifies an entire block of edges at each step of the chain to guarantee that the visited graphs always belong to the space of block structured graphs $\mathcal{B}$. By doing so, the search is limited to the subset of graphs whose structure is consistent with $V_B$. 
Moreover, exploiting a  trans-dimensional version of the Exchange algorithm \citep{murray2012mcmc}, 
our algorithm  
avoids any calculation of the G-Wishart normalizing constant.

We denote the current state of the chain by $\left(\bm{K}^{[s]}, G^{[s]}\right)$, where $\bm{K}^{[s]}\in \mathbb{P}_{G^{[s]}}$ and $G^{[s]}\in \mathcal{B}$. Since the graph is constraining the support of the precision matrix, the Reversible Jump technique is needed to handle trans-dimensional moves due to a different number of unknown entries of the precision matrix in subsequent iterations.  
In the first step of the algorithm, the state $(\bm{K}', G')$ is proposed, and the graph $G'$ is accepted or rejected. 
It consists of three parts: (i) a new  graph $G'$ is proposed in the neighborhood of the multigraph representative $G^{[s]}_B$ of the current graph $G^{[s]}$; (ii) a matrix $\bmK'$ compatible with the constraints imposed by $G'$ is constructed; (iii) the acceptance-rejection probability is computed by exploiting a modified Exchange algorithm \citep{murray2012mcmc}.
Note that, in part (ii) the proposed matrix $\bmK'$  must be guaranteed to be a precision matrix.
Differently from \citet{giudici1999decomposable}, who proposed a Reversible Jump sampler that limits itself to visit decomposable graphs and requires checking $\bmK'$ positive definiteness, we follow \citet{dobra2011bayesian} and \citet{alexlenkoski2013} and adopt a reparametrization 
based the Cholesky decomposition, so that $\bmK'$ is positive definite by construction.
The algorithm requires a double reversible move, leading to a Double Reversible Jump sampling strategy. 
In the following, each of these three parts is described. 

\subsubsection{(i) Proposing a new graph $G'$}
\label{section:graph}
A common feature of existing MCMC methods for graphical models is to build Markov chains such that the proposed graph $G'=(V,E')$ belongs to the one-edge-away neighborhood of $G$, which is defined as
\begin{equation}
    nbd_{p}(G) := nbd_{p}^{+}(G) \cup nbd_{p}^{-}(G), 
    \label{eqn:nbd_p}
\end{equation}
where $nbd_{p}^{+}(G)$ and $nbd_{p}^{-}(G)$ are the sets of undirected graphs having $p$ nodes that can be obtained by adding or removing an edge to $G\in\mathcal{G}$, respectively.
A step in an MCMC algorithm that selects $G'\in nbd_{p}(G^{[s]})$ is said to be a local move. The proposed BDRJ approach is innovative because it proposes moves that modify an entire block of edges instead of just a single one. 
In other words, our moves are local in the space $\mathcal{G}_{B}$ but not in the space $\mathcal{G}$.

Our procedure leverages the definition of the  map $\rho$ and generalizes standard graphical modeling tools to the space $\CcB$.
Hence, suppose $G^{[s]}\in \mathcal{B}$ and we aim to construct a new graph $G'\in \mathcal{B}$. Firstly, we map the current graph into its multigraph representative $G^{[s]}_{B} \in \mathcal{G}_{B}$, where $G_{B}^{[s]} = \rho^{-1}\left(G^{[s]}\right)$. Then, with probability $\alpha_G$, we add a new edge or, with probability $(1-\alpha_G)$, we remove one of its existing edges. Namely, the new multigraph representation $G'_B \in \CcG_B$ is drawn from
\begin{equation}
	q\left(G'_B\lvert G^{[s]}\right) = \alpha_G \tilda \text{Unif}\left(
	nbd^{\mathcal{B},+}_{M}\left( \rho^{-1}\left(G^{[s]}\right)\right)
	\right) + 
	(1-\alpha_G)\tilda\text{Unif}\left(
	nbd^{\mathcal{B},-}_{M}\left(\rho^{-1}\left(G^{[s]}\right)\right)
	\right),
	\label{eqn:proposal_G}
\end{equation}
where, similarly to Equation  \eqref{eqn:nbd_p}, $nbd^{\mathcal{B}}_{M}\left(G^{[s]}_{B}\right)$ is the one-edge-away neighborhood of $G_{B}^{[s]} 
$ with respect to the space of multigraphs $\mathcal{G}_{B}$. 
Finally, $\rho$ is applied  again to map the resulting multigraph back in $\mathcal{B}$ to obtain $G'$, i.e., we set $G' = \rho\left(G'_B\right)$.

If $\alpha_{G}=0.5$, Equation  \eqref{eqn:proposal_G} gives equal probability to addition and deletion moves. To lighten the notation, we always refer to this case.
The proposal distribution in Equation  \eqref{eqn:proposal_G} is preferred
over choosing uniformly in the whole neighborhood as in \citet{madigan1995bayesian}. Indeed, \citet{dobra2011bayesian} noticed that in a simple uniform edge sampling, the probability of proposing a move that adds (or deletes) an edge is too small if the current graph has a very large (or small) number of edges. Therefore, Equation \eqref{eqn:proposal_G} guarantees a better mixing in the resulting Markov chain.
Furthermore, Equation \eqref{eqn:proposal_G} reveals how the multigraph representation enables us to use standard tools of structural learning in the space $\mathcal{G}_{B}$ to get a non-standard proposal in the original space $\mathcal{G}$.

\subsubsection{(ii) Constructing the precision matrix $\bmK'\mid G'$}
\label{section:proposing_K}

Once the graph is selected, we need to specify a method to construct a proposed precision matrix $\bm{K}'$ that satisfies the constraints imposed by the new graph $G'$.
In principle, the method of \citet{wangh.lis.z.2012}, based on the partial analytical structure of the $\operatorname{G-Wishart}$, appears to be an efficient choice. However, this solution strongly relies on the possibility of writing down an explicit formula of the full conditional distribution of the elements of $\bm{K}$. Such a result, presented in \citet{roveratoa.2002}, can be handled in practice only if one edge of the graph is modified at each step of the algorithm. Instead, the proposal distribution presented in \Cref{section:graph} modifies an arbitrary number of edges. 
Differently, we build on a generalization of the Reversible Jump mechanism of \citet{alexlenkoski2013} and exploit the Cholesky decomposition  of $\bm{K}^{[s]}$  to guarantee the positive definiteness of $\bm{K}'$ and the zero constraints imposed by $G'$.

Indeed, $\bm{K}^{[s]}\in \mathbb{P}_{G^{[s]}}$ implies that it is possible to compute its Cholesky decomposition, $\bm{K}^{[s]}=\left(\bm{\Phi}^{[s]}\right)^{\top}\bm{\Phi}^{[s]}$, where $\bm{\Phi}^{[s]}$ is an upper triangular matrix. 
This is appealing because the zero constraints imposed by $G^{[s]}$ on the off-diagonal elements of $\bm{K}^{[s]}$ induce a precise structure and properties on $\bm{\Phi}^{[s]}$. 
In particular,
let $\nu\left(G^{[s]}\right) = \left\{(i,j)\tilda\lvert\tilda i,j\in V, i=j \lor (i,j)\in E^{[s]}\right\}$ be the set of edges belonging to $G^{[s]}$ plus the diagonal entries of its adjacency matrix. Hence,  $\bm{\Phi}^{\nu\left(G^{[s]}\right)}=\left\{\bmPhi_{ij} \tilda\lvert\tilda i,j \in \nu\left(G^{[s]}\right)\right\}$ is said to be the set of \textit{free elements} of $\bm{\Phi}^{[s]}$. The remaining entries
are uniquely determined through the completion operation \citep[Proposition  2]{a.ataykayish.massam2005} as a function of the free elements.
We  refer to these elements as the \textit{non-free elements}. See \cite{roveratoa.2002} and \cite{a.ataykayish.massam2005} for an exhaustive overview.

Suppose that the proposed graph $G'$ is obtained from Equation \eqref{eqn:proposal_G} by adding the edge $(l,m)$ to the  multigraph representation of $G^{[s]}$. The set of edges that are changing in $\mathcal{G}$ is then $L=\left\{(i,j)\tilda\lvert\tilda i,j\in V, i < j, (i,j)\in E', (i,j)\notin E^{[s]}\right\}$. The cardinality $\lvert L \lvert$ is arbitrary and, in general, greater than one. We call $V(L)=B_{l}\cup B_{m}$ the set of the vertices involved in the change. 
Note that $\nu(G')=\nu\left(G^{[s]}\right)\cup L$. Our solution to define the new free elements is to maintain the same value for all the ones that are not involved in the change and to set the new ones by perturbing the current, non-free elements, independently and  with constant variance $\sigma_{g}^{2}$. Namely, draw $\eta_{h}\mytilde{ind}\mbox{N}\left(\bmPhi_{h}^{[s]}, \sigma^{2}_{g}\right)$ and set $\bmPhi'_{h}=\eta_{h}$ for each $h\in L$. 
Then,  all non-free elements of $\bmPhi'$ are derived through completion operation \citep{a.ataykayish.massam2005} and the proposed precision matrix $\bm{K}'= (\bm{\Phi}')^{\top}\bm{\Phi}'$ is then obtained. Note that, we are generating a random variable $\bm{\eta}$ of length $|L|$ that matches the dimension gap between $\bm{K}^{[s]}$ and $\bm{K}'$.
In case of dimension reduction, say $\bmK\ats\rightarrow \bmK'$, the move is deterministic since it is defined in terms of the opposite move $\bmK'\rightarrow \bmK\ats$, where the extra elements do not need to be sampled. 
Then, the acceptance probability is the reciprocal acceptance probability of the corresponding increasing move.

\subsubsection{(iii) Computing the acceptance-rejection probability}
\label{section:step}
Finally, we frame the previous mechanism in the Exchange algorithm  paradigm \citep{murray2012mcmc} to eliminate the presence of the $\operatorname{G-Wishart}$ normalizing constants. 
Specifically, we employ the Double Reversible Jump procedure \citep{alexlenkoski2013}, which is the trans-dimensional equivalent of the Exchange algorithm. 

Let  $\bmwidetW$ be a latent $p \times p$ symmetric and positive definite matrix   that is Markov with respect to $G'$, i.e., $\bmwidetW\in\P_{G'}$. The matrix $\widetilde{\bm{W}}$ is sampled from a $\mbox{G-Wishart}(b,D)$ distribution using the exact sampler of \citet{alexlenkoski2013}. 
The BDRJ 
considers switching between 
 $   \left(\bm{K}\ats,G\ats,\widetilde{\bm{W}},G'\right)$
to the alternative 
 $  \left(\bm{K}',G',\bm{W}^{0},G\ats\right)$, with $\bmW^0\in\P_{G\ats}$, 
by performing two reversible jump moves: (i) a dimension increasing jump from $\left(\bm{K}\ats,G\ats\right)$ to $\left(\bm{K}',G'\right)$ according to the posterior parameters $b+n$ and $D+U$ of the G-Wishart distribution; (ii) a dimension decreasing jump from $\left(\widetilde{\bm{W}},G'\right)$ to $\left(\bm{W}^{0},G\ats\right)$ according to the prior parameters $b$ and $D$ of the G-Wishart distribution. Thus, the augmented target  is the joint distribution $p\left(\bmK\ats,G\ats,\bmwidetW,G'\mid\mathbf{y}\right) $ and the proposed graph $G'$ is accepted with probability $min(1, R^+)$, with 
\begin{equation}
\begin{split}
    R^+ =
    &
     \tilda
    \frac{p\left(\bmK',G',\bmW^0,G\ats \mid \mathbf{y}\right)}{p\left(\bmK\ats,G\ats,\bmwidetW,G'\mid \mathbf{y}\right)} \tilda
    \tilda      \frac{q\left(\bmK'\mid\bmK\ats\right)}{q\left(\bmW^0\mid\bmwidetW\right)}
    \frac{\mathcal{J}\left(\bmK', \bmW^0\right)} {\mathcal{J}\left(\bmK\ats, \bmwidetW\right)}
    \\
    = & \tilda
\frac{p\left(\mathbf{y}\mid\bmK',G'\right)}{p\left(\mathbf{y}\mid\bmK\ats,G\ats\right)}
    \frac{p\left(\bmK'\mid G'\right)}{p\left(\bmK\mid G\ats\right)}
    \frac{p\left( \bmW^0\mid G\ats\right)}{p\left(\bmwidetW\mid G' \right)}
            \frac{q\left(G\ats\mid G'\right)}  {q\left(G'\mid G\ats\right)} 
     \frac{\pi(G')}{\pi\left(G\ats\right)} \\
     & \times 
    \frac{q\left(\bmK'\mid\bmK\ats\right)}{q\left(\bmW^0\mid\bmwidetW\right)}
   \frac{\mathcal{J}\left(\bmK', \bmW^0\right)} {\mathcal{J}\left(\bmK\ats, \bmwidetW\right)}
       ,
    \label{eqn:BDRJ1}
    \end{split}
\end{equation}
where $q(\cdot, \cdot)$ 
denotes the density of the proposal distribution, and $\mathcal{J}(\cdot, \cdot)$ denotes the Jacobian of the transformations involved in the reversible moves, as  detailed in the online Appendix. 
Note that the normalizing constant ratio in the acceptance-rejection rate in Equation \eqref{eqn:BDRJ1} cancels out. 
The MCMC algorithm is then completed with a second step consisting in sampling the precision matrix $\bmK^{[s+1]}$ from its full conditional distribution, i.e., $\bm{K}^{[s+1]} \mid G^{[s]}, \mathbf{y} \sim  \operatorname{G-Wishart}(b+n,D+U)$.
The resulting algorithm is summarized in Algorithm 1 appearing in the online Appendix, together with the details on the derivation of the acceptance-rejection probability in Equation \eqref{eqn:BDRJ1}. The \texttt{R} package \texttt{BGSL}, implementing the BDRJ algorithm, is available at \url{github.com/alessandrocolombi/BGSL}.

Our proposed method BDRJ borrows the skeleton of the  Double Reversible Jump but modifies the proposal distribution to guarantee that $G'\in\mathcal{B}$ and, as a consequence, that multiple elements of the precision matrix are updated accordingly. 
Since only proposal distributions and priors have been changed, we still have a valid MCMC scheme that can now infer relationships in block structured graphs. 

\subsection{Posterior inference}
\label{section:posterior_inference}
Posterior inference of  the graph has to be performed with some care. In the ideal case, we would like to approximate its posterior distribution with the relative frequency of each sampled graph.
Then, one way of providing a pointwise estimate of the graph structure is to use the maximum a posteriori strategy, which represents the mode of the posterior distribution.
As noticed by \citet{jonesb.carvalhoc.dobraa.hansc.carterc.westm.2005}, for problems with even a moderate number of nodes $p$, the space to be explored is so large that the graph frequency can not be viewed as a good estimate of its posterior probability because each particular graph may be encountered only a few times in the  MCMC sampling \citep{peterson.stingo.vannucci2015}.  

A more practical and stable solution is instead to estimate the posterior edges inclusion marginally. 
Let $S$ be the size of the MCMC output, then the posterior inclusion probabilities are estimated as
\begin{equation}
    \hat{p}_{ij} = \frac{1}{S}~\sum_{s=1}^{S}\bbone\left((i,j) \in E^{[s]}\right),
    \label{eqn:plinks}
\end{equation}
where $\bbone\left((i,j) \in E^{[s]}\right)$ is the indicator function representing the inclusion of the edge between nodes $i$ and $j$ in the graph $G^{[s]}=\left(V,E^{[s]}\right)$ visited during the $s$-th iteration.
We call $\widehat{\mathbf{P}}$ the upper triangular matrix having elements $\hat{p}_{ij}$, for $i=1, \dots, p$ and $j=i, \dots, p$, that are the proportion of MCMC iterations, after the burn-in, in which the edge $(i,j)$ has been selected to be part of the graph. 
Since $\widehat{\mathbf{P}}$ contains the posterior probabilities of edge inclusion, the matrix represents the uncertainty of including or not an edge in the graph.

Pointwise graphical estimate $\widehat{G}$ is carried out by selecting all edges whose posterior inclusion probability in Equation  \eqref{eqn:plinks} exceeds a given threshold $\tau$. A possible choice is $\tau=0.5$, in analogy with the median probability model of \citet{barbieri2004}, originally proposed in the linear regression setting. A second possibility is based on the Bayesian False Discovery  Rate (BFDR; \citealt{muller2007,peterson.stingo.vannucci2015})
\begin{equation}
    \mbox{BFDR} = \dfrac{\sum_{i < j}(1 - \hat{p}_{ij})\bbone\left(\hat{p}_{ij} \geq s\right)}{\sum_{i < j}\bbone\left(\hat{p}_{ij} \geq s\right)},
    \label{eqn:BFDR}
\end{equation}
where  $\tau$ is selected so that BFDR is below $ 0.05$.

For what concerns the precision matrix, we average over the MCMC samples \citep{cremaschi2019,wangh.lis.z.2012} to obtain the posterior mean
$\widehat{\bm{K}}$
\begin{equation}\nonumber
    \widehat{\bm{K}} = \frac{1}{S}\sum_{s=1}^{S}~\bm{K}^{[s]}.
    \label{eqn:BGSL_precision_estimate}
\end{equation}
Note that, even if each $\bm{K}^{[s]}$ has a block structure induced by $G^{[s]}$, $\widehat{\bm{K}}$ does not share the same structure of the selected graph $\widehat{G}$.

%% file: 4-SimulationStudy.tex
\section{Simulation study}
\label{section:simulation}
We carry out a simulation study to evaluate the ability of our methodology to recover the structure of the generating graph.
We compare our performance to the Birth and Death approach (BDgraph for short) proposed by \citet{mohammadi.wit2015} for Gaussian graphical models with a standard non-informative prior on the graph space.
We rely on the implementation provided by the corresponding \texttt{R} package \texttt{BDgraph} \citep{BDgraphpackage}.
In the latter, authors derived an efficient MCMC method where moves are decided according to specific birth and death transition kernels for the edges. Every proposed move is accepted, so the chain converges very quickly. Moreover, using an approximation of the G-Wishart normalizing constants ratio approximation 
\citep{mohammadi2017ratio} allows to speed up calculations.
In addition, we employ the proposed BDRJ algorithm assuming a partition of nodes where each node forms a single block; this boils down to the exact sampler called Double Reversible Jump (DRJ; \citealt{alexlenkoski2013}). Graph posterior estimates from both BDRJ, DRJ, and BDgraph approaches have been obtained by cutting the posterior probability of inclusion of each edge with the threshold chosen via the BDFR in Equation \eqref{eqn:BFDR}. 

We consider two different simulation scenarios to compare 
the ability of the aforementioned methods to learn the structure of conditional dependencies.
In the first experiment, we generate data using an underlying graph whose adjacency matrix has a block structure, i.e., with fixed groups of nodes. 
In the second experiment, we investigate the performance of the block graph prior model when the true graph has incomplete blocks and some isolated edges.

\subsection{Performance evaluation}
To assess the performance of recovering the graph structure, we compute the standardized Structural Hamming Distance (Std-SHD, \citealt{SHD}) from the underlying graph, which is, in the case of undirected graphs, equal to the number of wrongly estimated edges, standardized with respect to the number of all possible ones, i.e.,
\begin{equation}
    \text{Std-SHD}~=~\frac{\text{FP} + \text{FN}}{\binom{p}{2}},
    \label{eqn:std_SHD_def}
\end{equation}
where FP and FN are the number of false positives and false negatives, respectively.
Following 
\citet{osborne2021latent}, we also take in consideration the $\text{F}_{1}\text{-score}$, 
defined as
 \begin{equation}
     \text{F}_{1}\text{-score}~=~\frac{2\text{TP}}{2\text{TP} + \text{FP} + \text{FN}} ,
     \label{eqn:F1_score_def}
 \end{equation}
where TP is the number of true positives. 
Both indices lie between $0$ and $1$: for the $\text{Std-SHD}$ lower values are preferred ($0$ value stands for perfect match), while for $\text{F}_{1}\text{-score}$ higher values correspond to better performances ($1$ value stands for perfect match). 
The main difference between the two indices is that $\text{Std-SHD}$ equally weights errors due to false positiveness or negativeness while $\text{F}_{1}\text{-score}$ places higher importance on the number of correct discoveries that are the true positives. To visualize their difference, consider the following simple example: set the true graph to have a sparsity index equal to $0.1$ and consider a trivial estimator, i.e., the empty graph. The resulting $\text{Std-SHD}$ score would equal $0.1$, which seems reasonably good even if the estimated graph is not capturing any significant information. On the other hand, the $\text{F}_{1}\text{-score}$ score is not deceived as it would be equal to $0$.
In addition to the previous two indices, we compute the sensitivity index, i.e., $\text{TP}/(\text{TP}+\text{FN})$ and the  specificity index, i.e., $\text{TN}/(\text{TN}+\text{FP})$, where TN is the number of true negatives.

\subsection{Results}
\subsubsection{Experiment $1$ - Complete Blocks}
We set $n = 500$, $p=40$, and $M=p/2$ groups of equal size, which leads to off-diagonal blocks of size $2\times2$. The underlying blocked structure graphs have been randomly generated by sampling from $\pi(G\mid  \theta)$ with different sparsity indices $\theta$, uniformly distributed in the interval $\left[0.2,0.6\right]$.  
Given the graph, the true precision matrix has been sampled from a $\operatorname{G-Wishart}\left(3,\bm{I}_{p}\right)$. For this study, $\sigma^{2}_{g}$ was set equal to $0.5$, after a little tuning phase.
The MCMC sample comprises $400,000$ iterations plus $100,000$ extra iterations that were discarded as a burn-in period.
\begin{figure}
	\centering
	\includegraphics[width=0.235\linewidth]{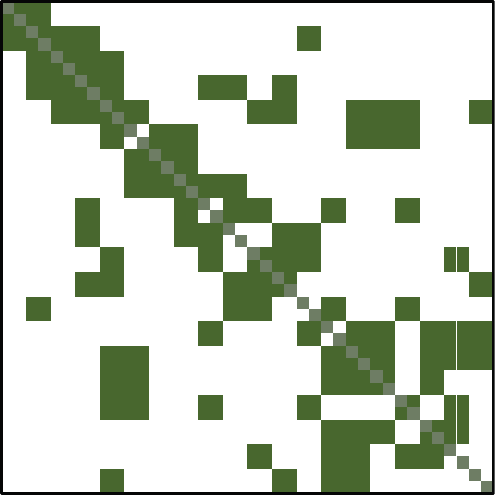}
	\hfill
	\includegraphics[width=0.235\linewidth]{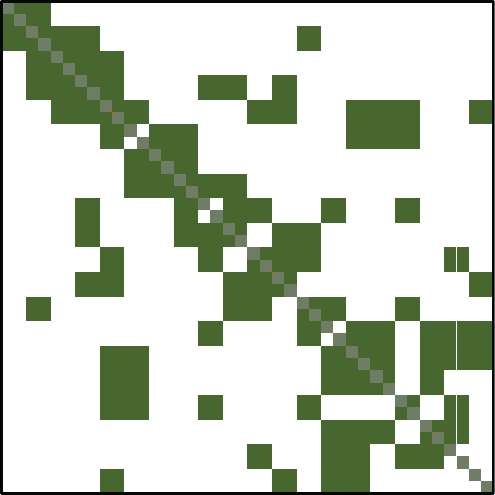}
	\hfill
	\includegraphics[width=0.235\linewidth]{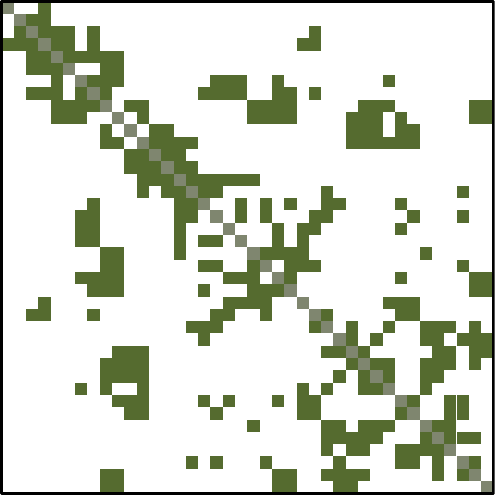}
	\hfill
	\includegraphics[width=0.235\linewidth]{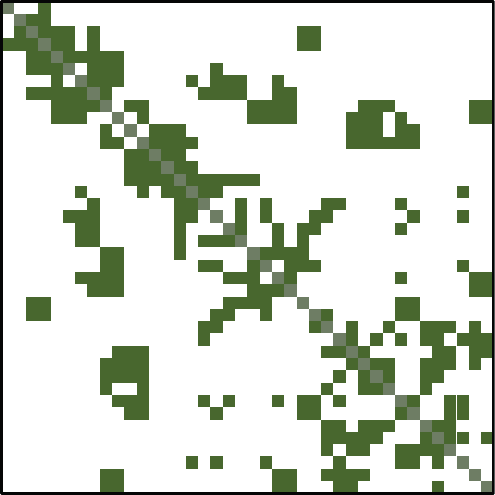}
	\caption[ Experiment $1$, estimated graphs]{ 
		From the left: the true graph (first panel) and the estimated graph obtained using BDRJ (second panel), DRJ (third panel), and BDgraph (fourth panel) for Experiment 1. Dark squares represent the included edges.}
	
	\label{fig:test_GGM}
\end{figure}
\Cref{fig:test_GGM} shows an example of the true graph and compares the final estimates of the forecited methods; green squares mean that there is an edge between the corresponding nodes.
The second, third, and fourth panels from the left display the graph estimated by BDRJ, DRJ, and by BDgraph, respectively.
Clearly, the visual inspection suggests that BDRJ provides a better estimate of the underlying graph than the competitors.

\begin{figure}[h!]
    \centering
	\includegraphics[width=0.32\linewidth]{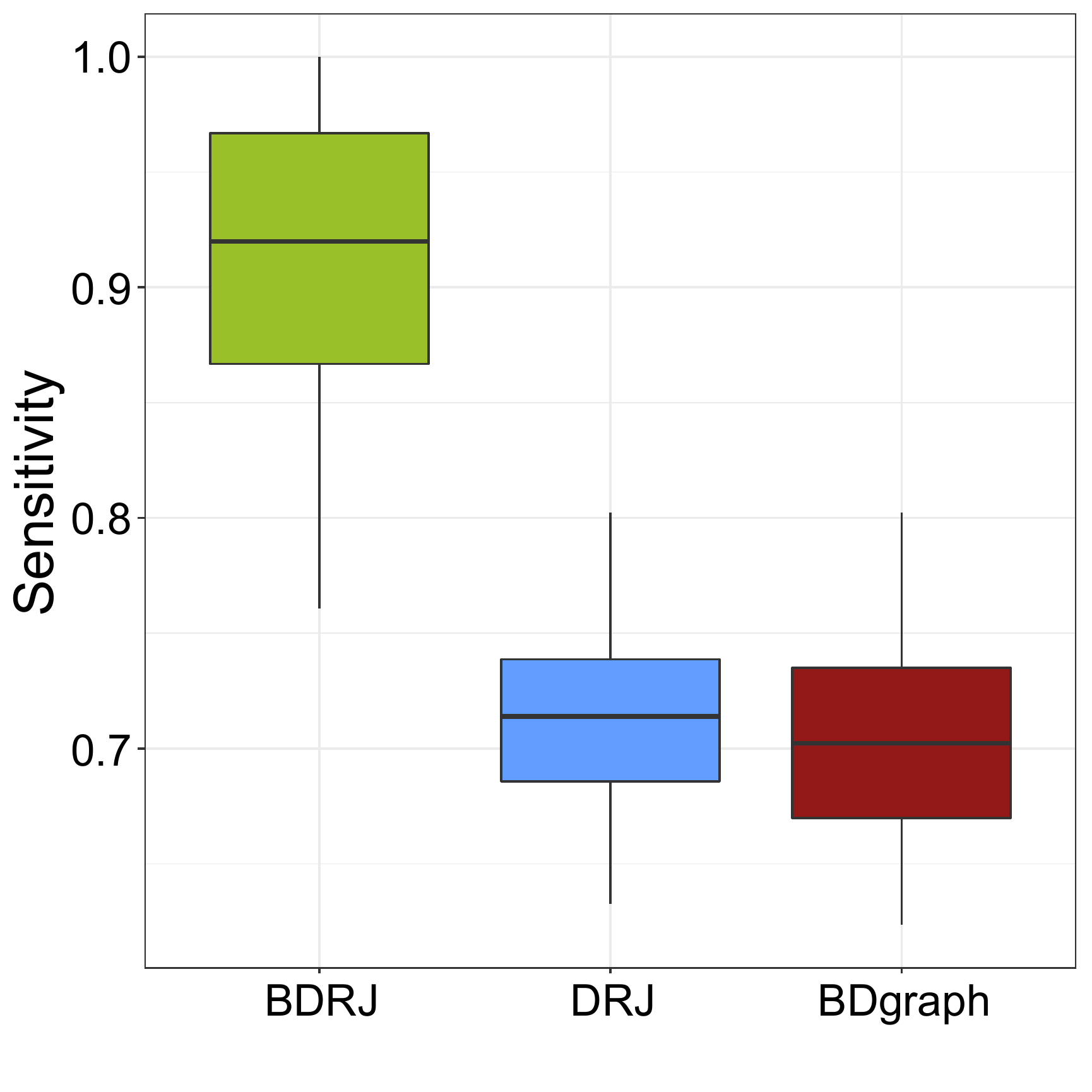}
	\includegraphics[width=0.32\linewidth]{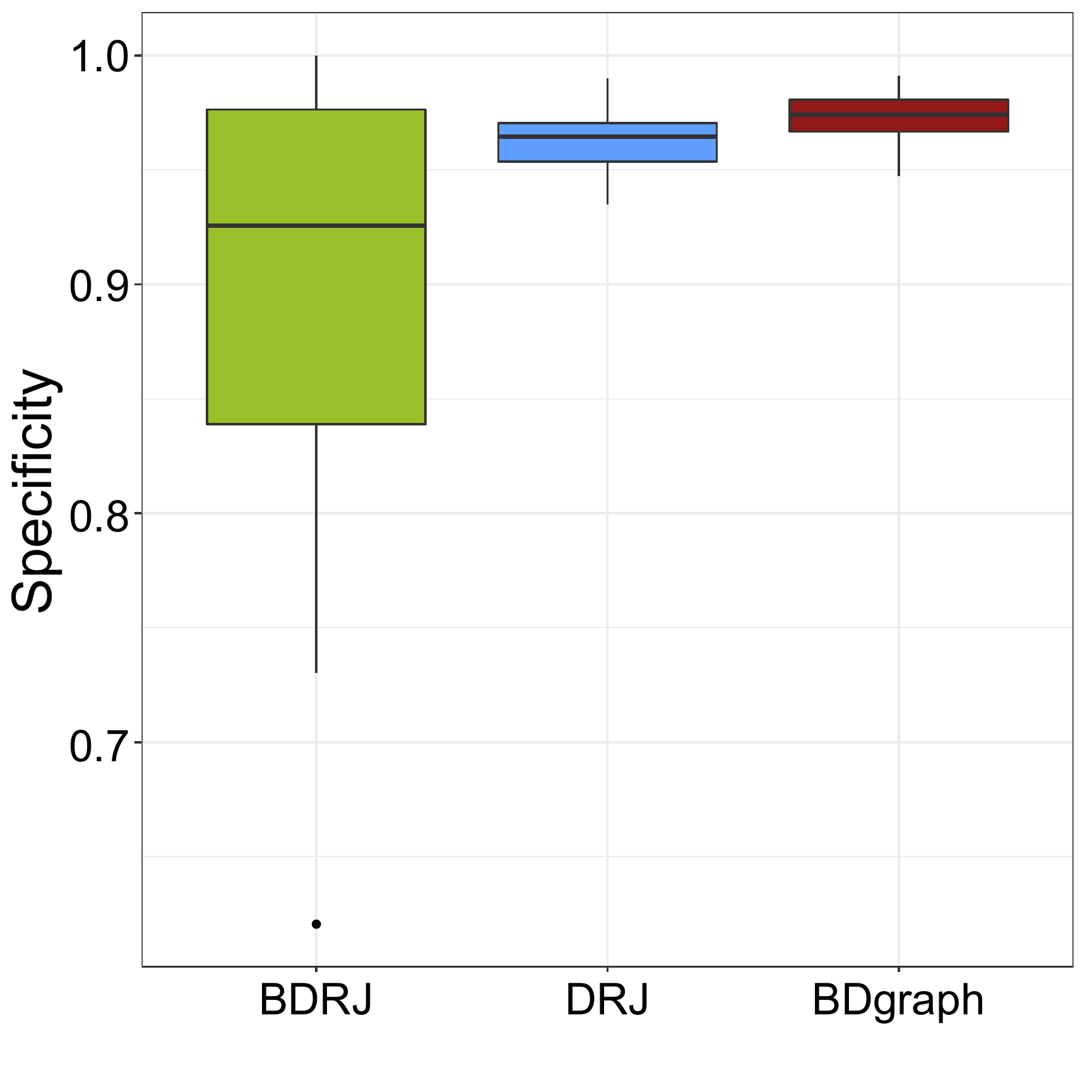}\\
	\includegraphics[width=0.32\linewidth]{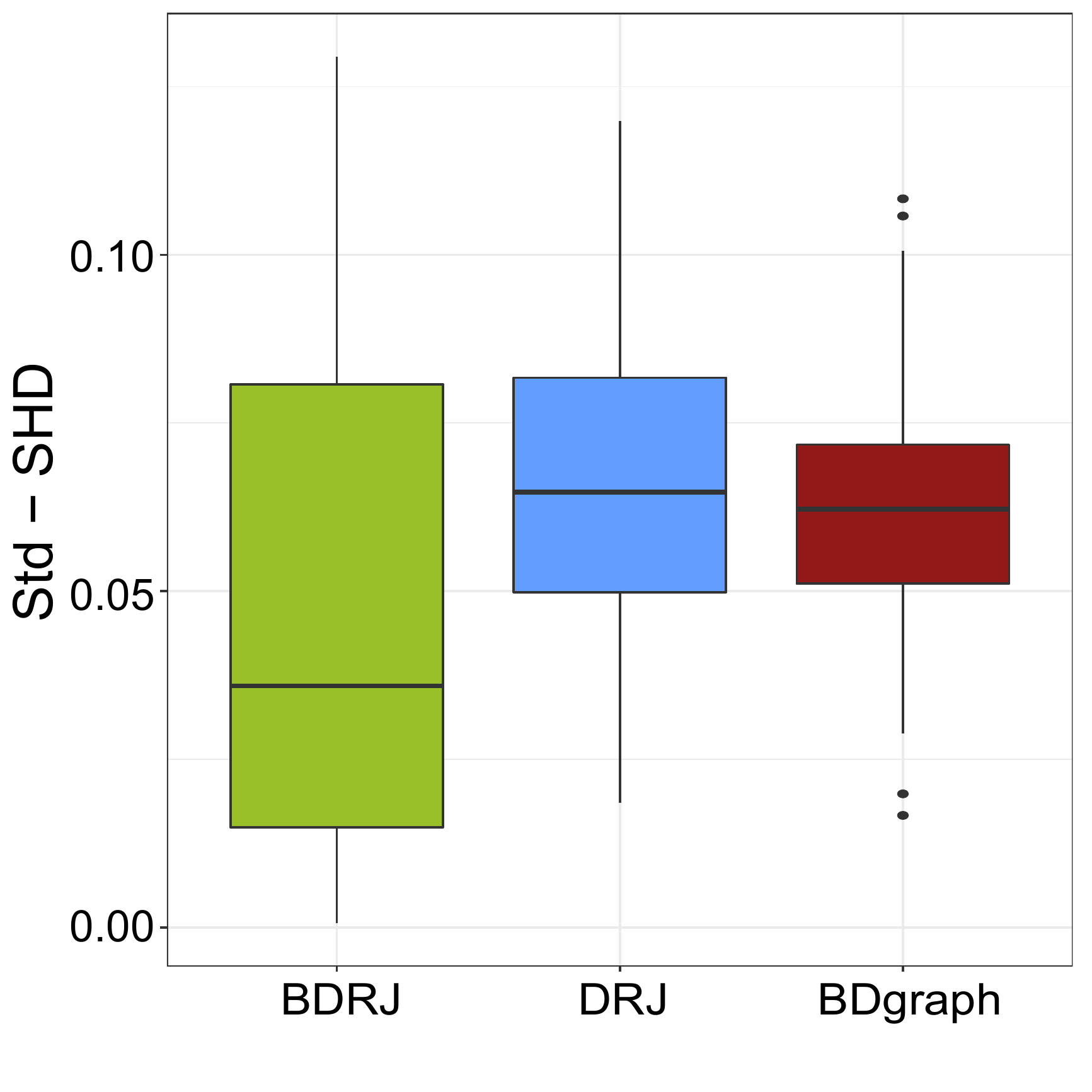}
	\includegraphics[width=0.32\linewidth]{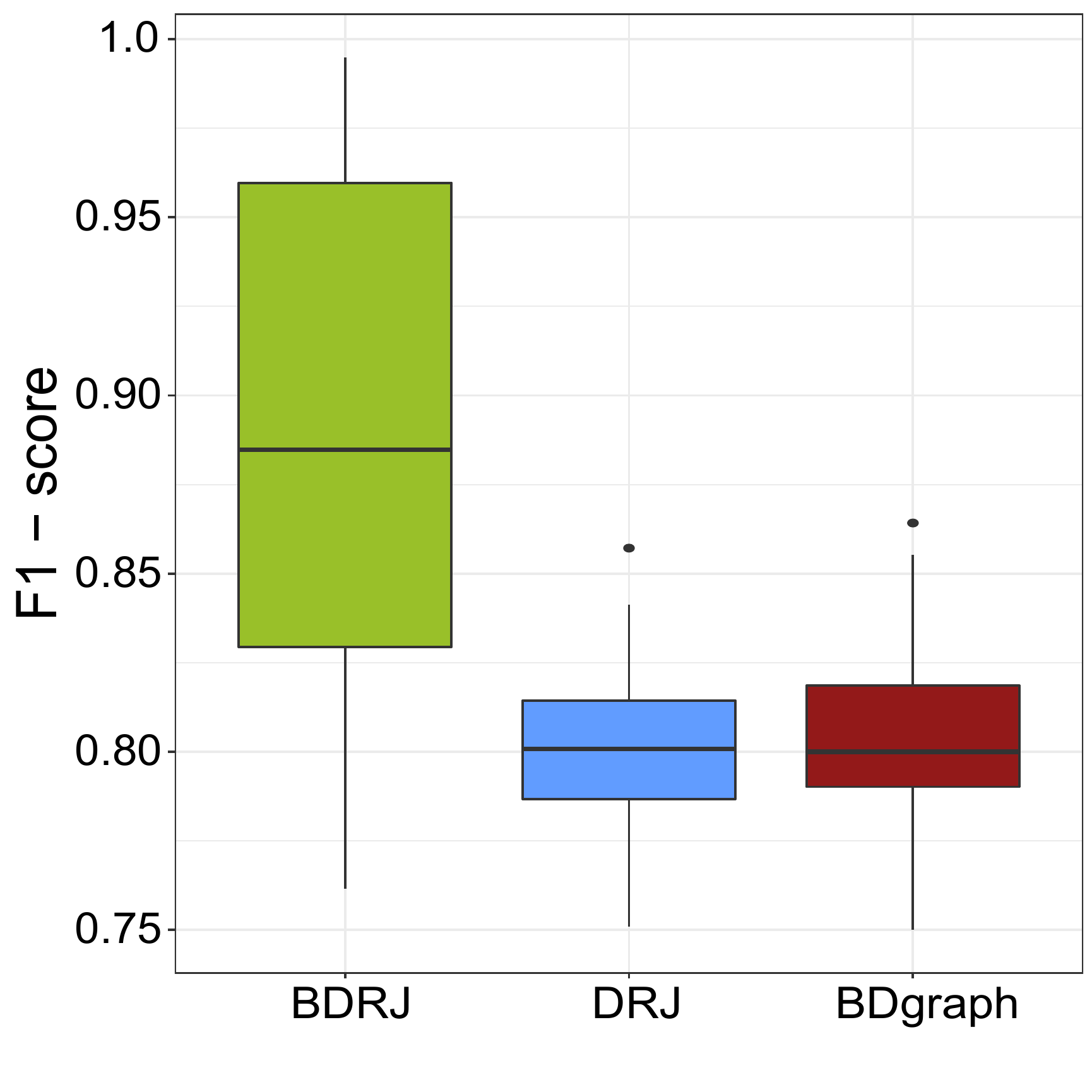}
	\caption[\texttt{GGMsampler}, experiment $1$, indices comparison]{  
	Boxplot of the
    sensitivity index (top-left panel),  specificity index (top-right panel),  Std-SHD (bottom-left panel), and F$_1$-score (bottom-right  panel) 
    over the simulated datasets of Experiment 1. 
	}
	\label{fig:GGM_p40_rep}
\end{figure}

\Cref{fig:GGM_p40_rep} shows the boxplot of  the sensitivity index, the specificity index, the
Std-SHD and the F$_{1}$-scores over 50 simulated datasets.  
The sensitivity and specificity indices of BDRJ are almost similar, being both centered around $0.9$. This denotes a good balance between including or not blocks of edges, without any preference between being conservative or not. Rather, the specificity of DRJ and BDgraph is close to one, i.e., a much higher value than the sensitivity index that is around $0.7$. Therefore, we conclude that DRJ and BDRJ tend to be more reliable in terms of discovered conditional independence relationships. 

We note that, overall, BDRJ outperforms the competitors in terms of both  the Std-SHD and the F1-score. 
The number of misclassified edges is rather low for BDRJ, with a median value of $\text{Std-SHD}$  equal to $0.0455$. 
Many true discoveries are achieved, and indeed the median $\text{F}_{1}\text{-score}$ under BDRJ is $0.885$. 
BDgraph and DRJ perform worse with respect to both indices; the medians Std-SHD are equal to $0.0622$ and $0.0647$ while the F$_{1}$-scores are both equal to $0.807$. 
The reason for these differences is that our approach takes advantage of the block structure of the true graph, which is not incorporated into the models used by the other methods. Instead, these methods attempt to estimate every possible link independently, leading to more errors in the final estimate and a less interpretable graph structure. It is difficult to explain why certain edges are missing within grouped structures.

\subsubsection{Experiment 2 - Incomplete blocks} 
In this experiment, we analyze the performance of our model when the  underlying graph has incomplete blocks.
To simulate the data under this scenario, we first draw a block structured graph from $\pi(G\mid \theta)$, where $\theta=0.2$. Then, edges are removed within each block with a probability equal to $0.25$. By doing so, the block structure is incomplete but still recognizable. 
We set $n=500$, $p=30$,  and  $M=p/2$ groups of equal size; given the graph, the true precision matrix has been sampled from $\operatorname{G-Wishart}(3,\bm{I}_{p})$.

\begin{figure}
	\centering
	\includegraphics[width=0.235\linewidth]{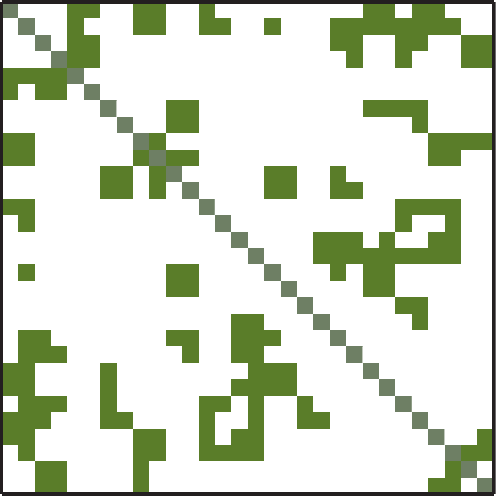}
	\hfill
	\includegraphics[width=0.235\linewidth]{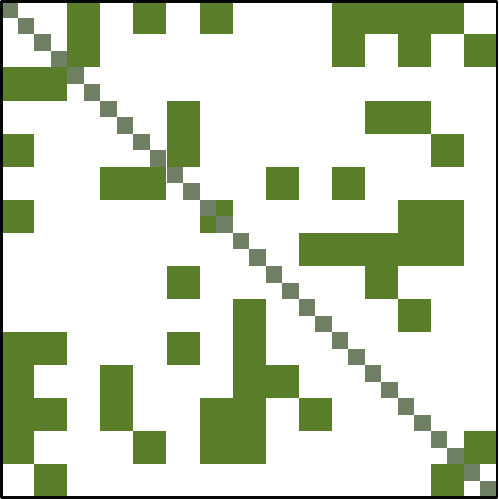}
	\hfill
	\includegraphics[width=0.235\linewidth]{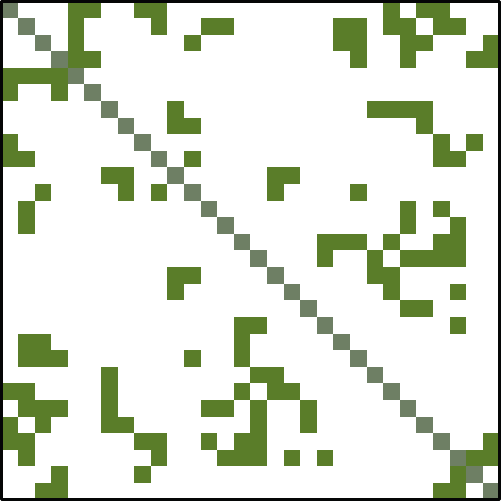}
	\hfill
	\includegraphics[width=0.235\linewidth]{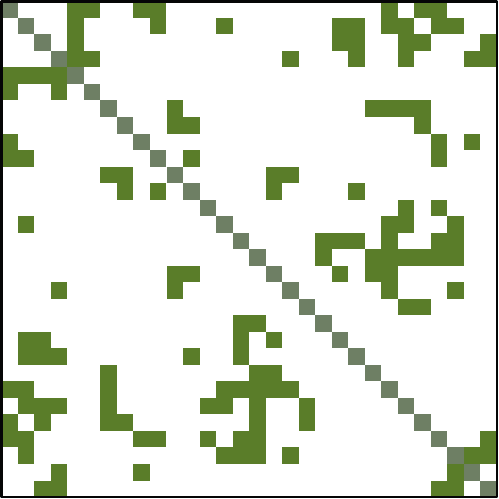}	
    \caption[\texttt{GGMsampler}, experiment $3$, estimated graphs]{
    From the left: the true graph (first panel) and the estimated graph obtained using BDRJ (second panel), DRJ (third panel), and BDgraph (fourth panel) for Experiment 2. Dark squares represent the included edges.
    }
    \label{fig:GGM_blocchi_mozzi_grafi}
\end{figure}

\Cref{fig:GGM_blocchi_mozzi_grafi} shows an example of the true graph (first panel from the left) and the estimated one using BDRJ (second panel), DRJ (third panel) and BDgraph (right panel). 
A simple visual inspection of the figure suggests that our approach tends to include incomplete blocks rather than discard them, leading to some false discoveries, as expected. 
On the other hand, both BDgraph and DRJ do not make any assumptions about the graph structure, so in principle, they should be able to recover the graph correctly. In practice, as soon as the dimension of the graph increases, this is unlikely to happen. Instead, the tendency is to be more conservative, which leads to fewer false discoveries. 

\begin{figure}[h!]
    \centering
	\includegraphics[width=0.32\linewidth]{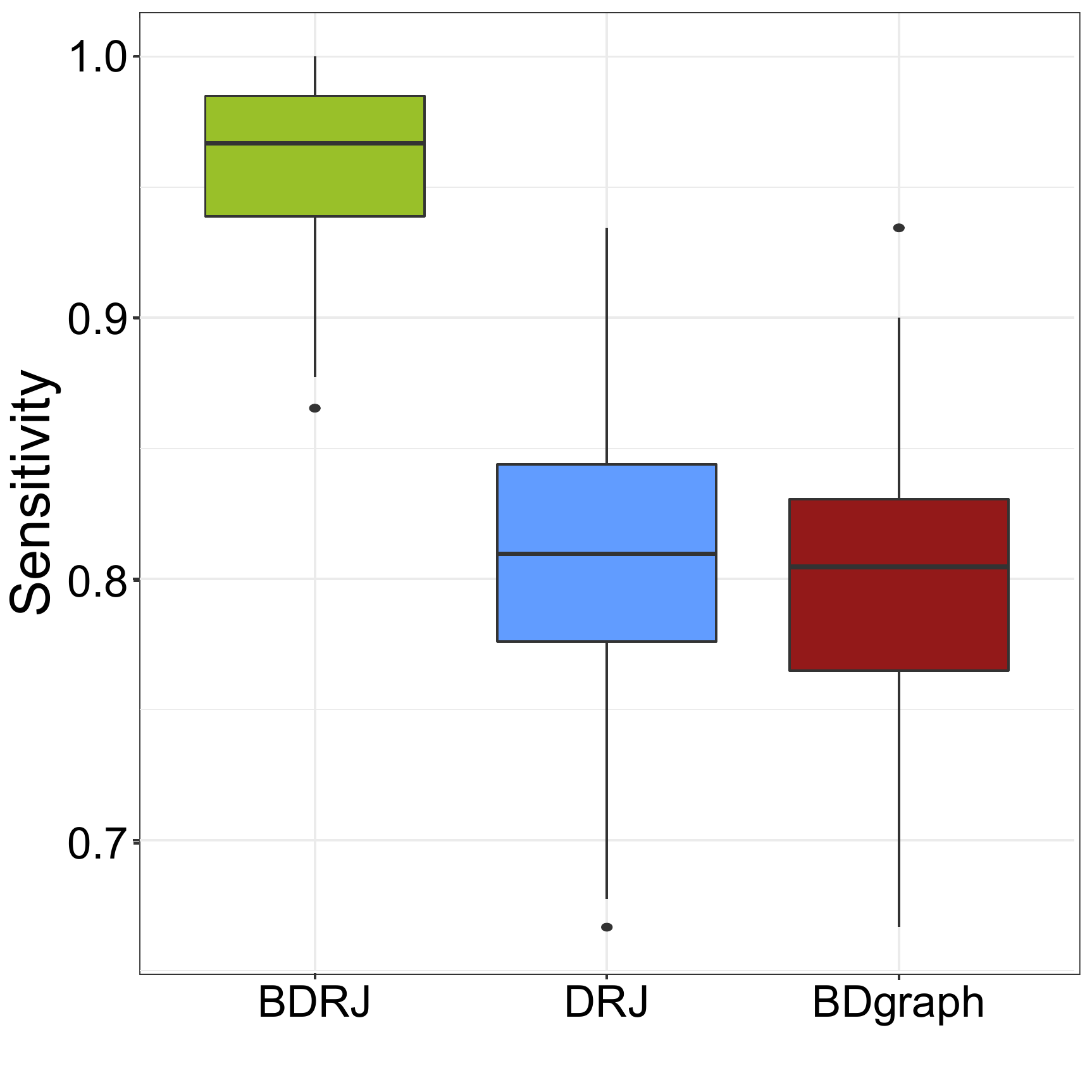}
	\includegraphics[width=0.32\linewidth]{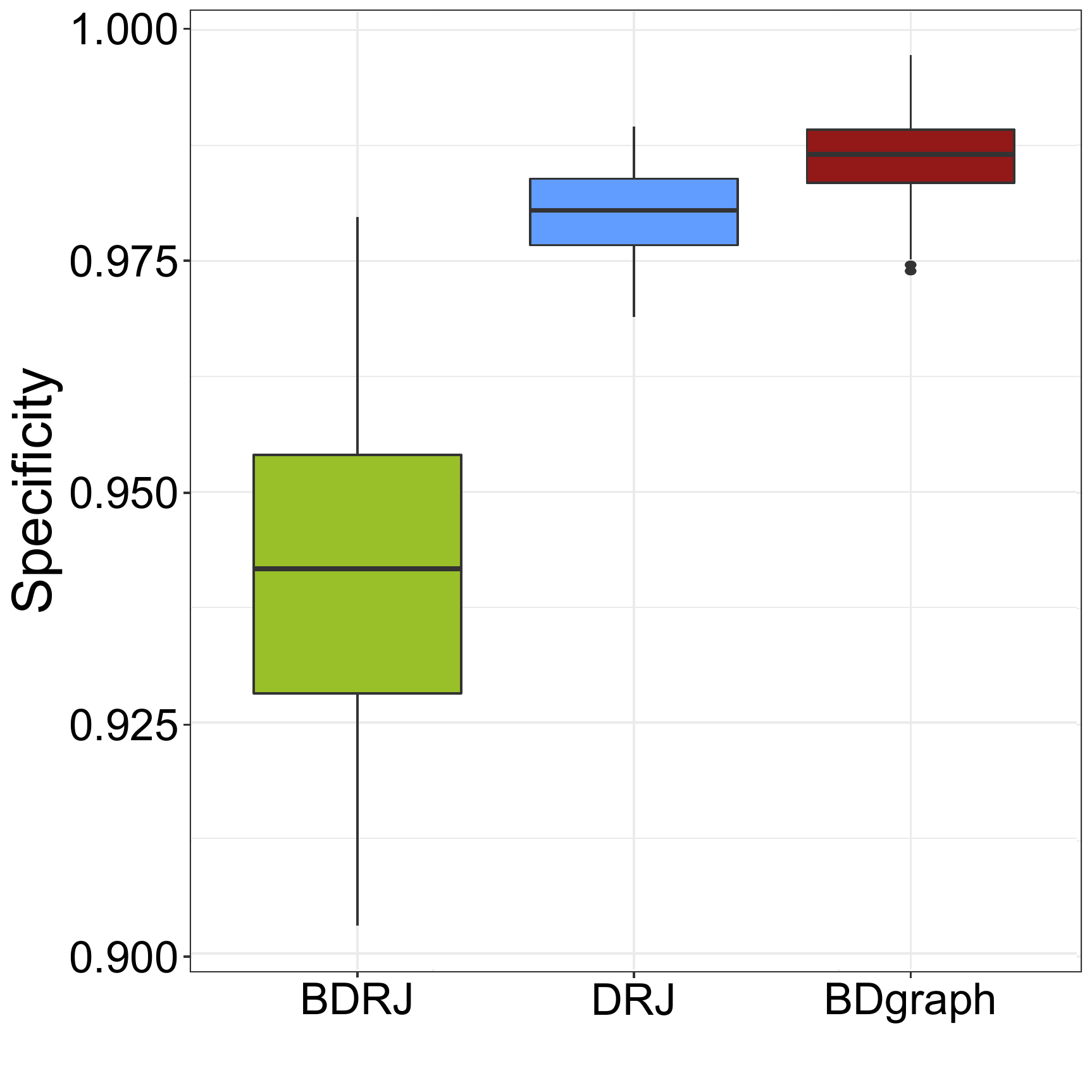}\\
	\includegraphics[width=0.32\linewidth]{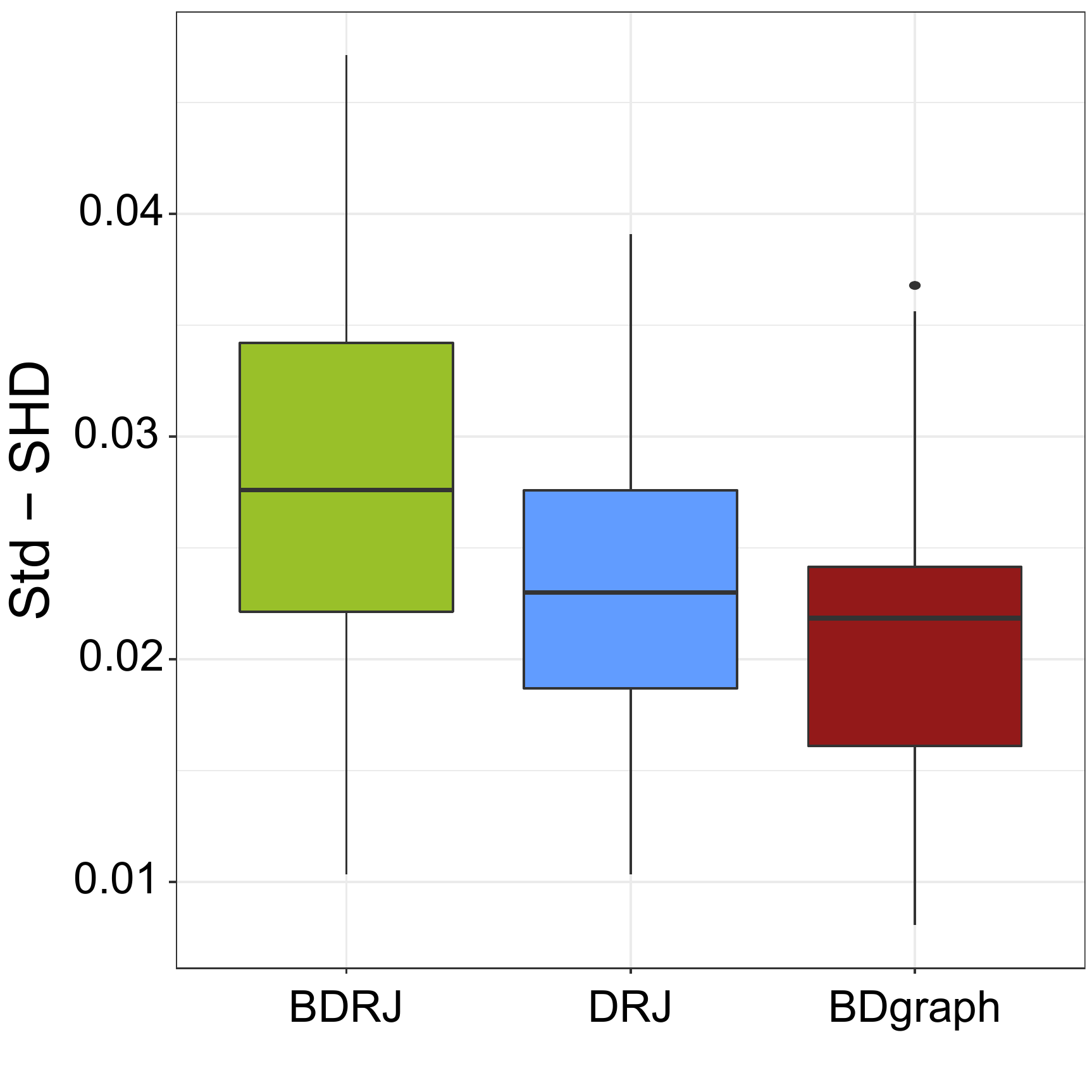}
	\includegraphics[width=0.32\linewidth]{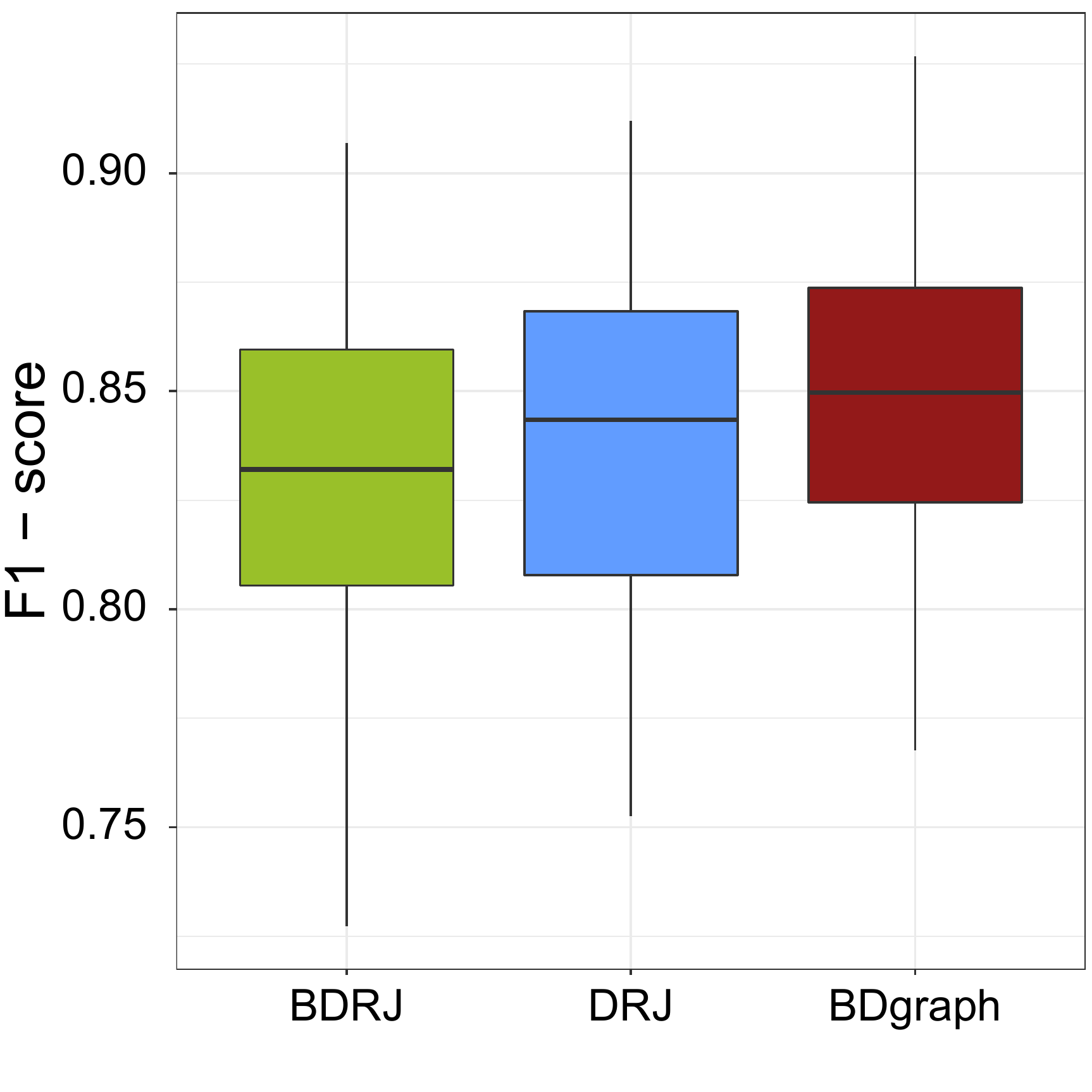}
	\caption[\texttt{GGMsampler}, experiment $3$, estimated graphs]{
	Boxplot of the
    sensitivity index (top-left panel),  specificity index (top-right panel),  Std-SHD (bottom-left panel), and F$_1$-score (bottom-right panel) 
    over the simulated datasets of Experiment 2. 
	}
	\label{fig:GGM_blocchi_mozzi_multiple_scores}
\end{figure}

To clarify, we report in \Cref{fig:GGM_blocchi_mozzi_multiple_scores} the boxplots of the sensitivity and specificity indexes computed over 50 simulated datasets. 
Our approach outperforms the competitors in terms of sensitivity
since it provides more true discoveries and fewer false negatives.
This means that it is unlikely that a missing edge is instead present in the underlying graph.
On the other hand, the BDgraph solution is preferable to BDRJ and DRJ in terms of specificity, i.e., an included edge likely represents   an actual connection in the underlying graph. 
As expected, DRJ and BDgraph show similar performances, still with slightly different behavior. As observed also in Experiment 1, the effect of the exact sampler used in DRJ seems to be an increase of sensitivity at the price of reduced specificity. 
The specificity and sensitivity indices provide a clear picture of the differences between BDRJ and the two competitor approaches,  
which is no longer true looking at the Std-SHD and the F$_1$-score values in \Cref{fig:GGM_blocchi_mozzi_multiple_scores}. The Std-SHD is slightly higher for BDRJ, coherently with the fact that the prior information assumed in this case is misspecified, but overall the difference between the three methods is limited.   
The F$_1$-score, in particular, is very similar for the three methods, meaning that the three approaches are almost equivalent in terms of misclassified edges. In other words, when dealing with a structured graph with incomplete blocks, the expected findings from a stochastic search in the complete or block graph space are comparable, but with the latter depicting more interpretable results.

%% file: 5-Applications.tex
\section{Analysis of fruit purees}
\label{section:fruit_purees}
\begin{figure}
    \centering
    \includegraphics[width=\linewidth]{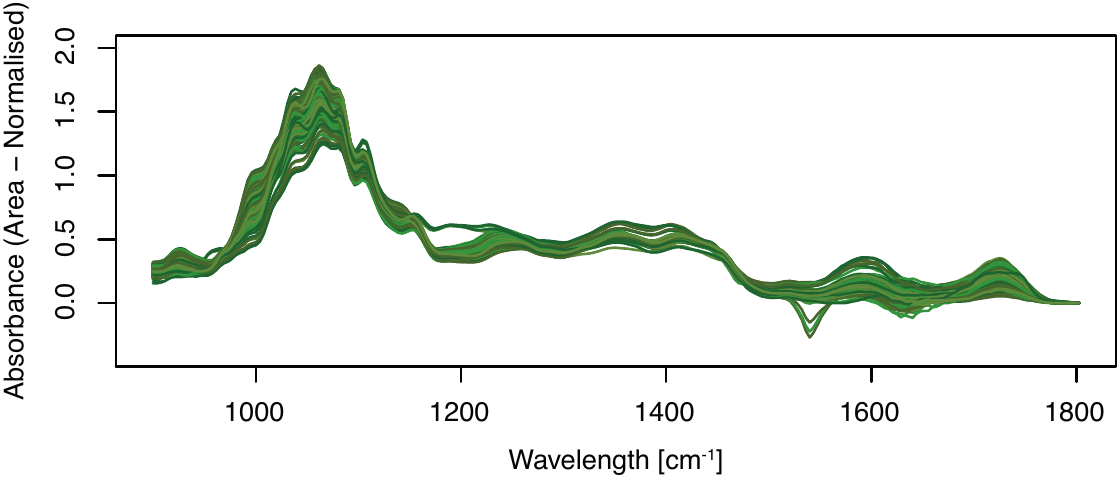}
    \caption[Fruit purees dataset]{Plot of 351 spectra of absorbance of pure fruit purees, measured in 235 different wavelengths of the middle-infrared spectra.}
    \label{fig:purees_true}
\end{figure}

We illustrate  with a real-world application how our model is able to exploit prior information to enrich the data analysis and provide more interpretable results. The motivating problem is the analysis of spectrometric data of fruit purees \citep{zheng2019spectra,waghmare2023}, which are publicly available at \url{https://data.mendeley.com/datasets/frrv2yd9rg}. See \cite{holland.kemsley.wilson1998} for an exhaustive description of the dataset. Our analysis focuses on $351$ absorbance spectra of strawberry purees, that are displayed in \Cref{fig:purees_true}.
Curves were measured on an equally spaced grid of $235$ different wavelengths, whose range is $\left[899,1802\right]~cm^{-1}$. The resulting spectra were then standardized with respect to the area under the curve so that their final range is $\left[-0.1,1.7\right]$.
The shape of the spectra is very similar for all curves: they all exhibit a well-recognizable peak around the wavelength $1000-1200~cm^{-1}$ and few secondary others, like the ones around $1600~cm^{-1}$ and $1700~cm^{-1}$.
The data were collected using infrared spectroscopy, which measures the interaction of infrared radiation with the matter by absorption, emission, or reflection. Such a technique is used to study and identify chemical substances or functional groups in solid, liquid, or gaseous forms. Indeed, from a chemical perspective, the spectrum can be seen as the identity card of the substances that are present within the compound. However, the spectrum of heterogeneous compounds may be difficult to analyze due to overlapping emissions of different substances interacting with one another.

The problem has already been addressed in \citet{codazzi2021gaussian}, where data were previously analyzed. From a mathematical point of view, a spectrum is a continuous function of the wavelength, and so the authors framed the problem within the functional data analysis setting. The classical smoothing strategy was enriched by placing a  Gaussian graphical model on the basis expansion coefficients, providing an estimate of their conditional independence structure. Since the elements of a B-Spline basis have compact support, the conditional independence structure is reflected on well-defined portions of the domain.
The Bayesian hierarchical formulation enables the borrowing of strength among different curves, and the graphical model allows sharing of information along different subintervals of the functional datum.
Finally, note that, in this application, the support of the spline basis functions coincides with spectrum bands. Therefore the problem of studying interactions between different substances simply translates into studying the dependencies between the basis expansion coefficients which can be read from the graph.

\citet{codazzi2021gaussian} used independent $\operatorname{Bernoulli}$ prior distributions on the graph edges and the BDgraph method to provide posterior inference on the graph and the precision matrix. As already discussed, this is a general non-informative setting that does not allow to include further prior information even when they are available. 
This is one of those situations: for infrared spectrometric data, peaks of the signals are associated with the vibrational modes of the different molecules present in the substance \citep{atkins2013elements}. This is why the signals can be decomposed into different parts, corresponding to the peaks observed along the domain. 
As a matter of fact, domain experts identified nine intervals of the spectrum of chemical interest associated with the most significant peaks of the signal \citep{defernez1995use}.

\begin{figure}[!ht]
    \centering    
        \includegraphics[width=0.4\textwidth]{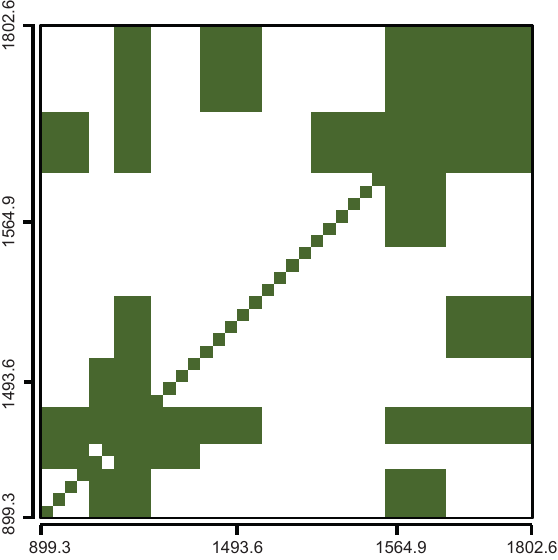}
        \hfill    \includegraphics[width=0.55\textwidth]{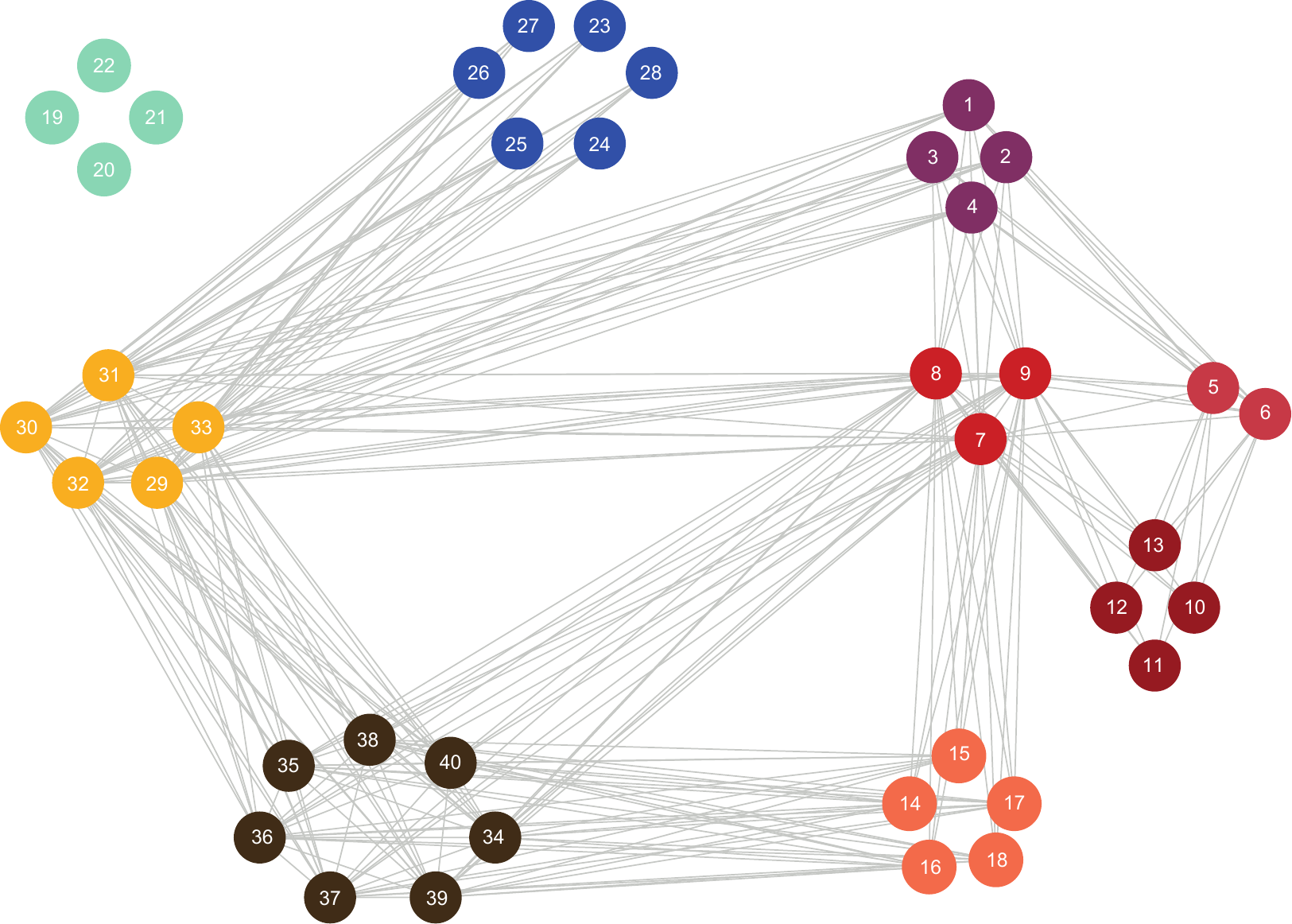}\\
    \vspace{4mm}
    \includegraphics[width=1\textwidth]{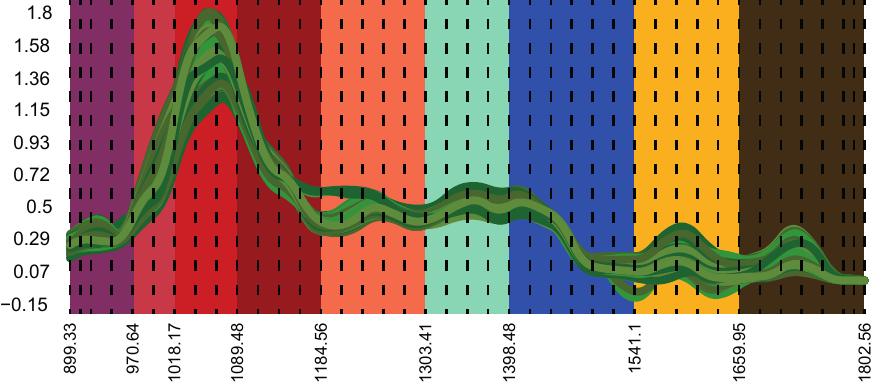}
   \caption{Top panels: a posterior graph estimate obtained using the BFDR criterion. Nodes are colored according to different portions of the spectra highlighted in the bottom panel, where different colors represent the nine groups.}
   \label{fig:purees_BDRJ} 
\end{figure}

The partition can be visualized in \Cref{fig:purees_BDRJ}, where different colors highlight the nine groups. Clearly, the nodes representing the basis expansion coefficients are ordered, so the groups are contiguous in the functional domain. The figure also shows the support of $p=40$ spline functions; the number of basis was chosen following \citet{codazzi2021gaussian}, as a trade-off between good fitting of the smoothed curves and limited computational burden.

Let $\bm{y}_t = \left(y_t(s_1), \dots, y_t(s_r) \right)^\top$ be the absorbance spectrum at all observed wavelengths of curve $t$, with $t=1, \dots, T=351$. We employ the block structured Gaussian graphical model described in Section \ref{section:blockprior} to smooth the functional data and accommodate prior knowledge on the subintervals of the spectrum. In particular, we assume $\pi_{B}\left(G_B\mid \theta \right)=\theta^{\lvert E_B\lvert}(1 - \theta)^{\binom{M}{2} - \lvert E_B \lvert}$, where $\theta = 2/(M-1)=0.25$, that is the choice suggested by \cite{jonesb.carvalhoc.dobraa.hansc.carterc.westm.2005} but taking into account that the number of nodes in the multigraph space is $M=9$, not $p=40$. All the other prior distributions and hyperparameters are set as in \citet{codazzi2021gaussian}. Summing up, the Bayesian hierarchical model is defined as follows:
\begin{nalign}
    \bm{y}_{t}~\lvert~\bm{\beta}_{t},~\tau^{2} &~\mytilde{ind}~\mbox{N}_{r}\left(\bm{\Omega}\bm{\beta}_{t},\tau^{2}\mathbf{I}_{r}\right),\\
	\bm{\beta}_{1},\dots,\bm{\beta}_{T}~\lvert~\bm{\mu},~\bm{K}&~\mytilde{iid}~\mbox{N}_{p}\left(\bm{\mu},\bm{K^{-1}}\right),\\
			\bm{K}~\lvert~ G &~\mytilde{~}~\operatorname{G-Wishart}\left(d,\bm{D}\right), \\
	G&~\mytilde{~}~\pi\left(G\right),\\
	\bm{\mu}&~\mytilde{~}~\mbox{N}_{p}\left(\bm{0},\sigma^{2}\mathbf{I}_{p}\right) \text{\ and}\\
		\tau^2 &~\mytilde{~}~ \operatorname{IG}\left(a,b\right),
	\label{fgm_model}
\end{nalign}
\noindent where the $lj$-th element of the matrix $\bm{\Omega}$ is the $j$-th basis function  evaluated at the $l$-th grid point $s_l$, $l=1, \dots, r=235$, $\mathbf{I}_r$ denotes the identity matrix of size $r$, $\pi(G)$ is defined in Equation \eqref{eqn:truncated} and $\operatorname{IG}(a,b)$ denotes the Inverse Gamma distribution with shape parameter $a$ and rate parameter $b$. 
Posterior inference of model in Equation \eqref{fgm_model} is obtained via the BDRJ algorithm run for $450,000$ iterations after a burn-in of $50,000$ and a thinning value of $25$. After some tuning, we set $\sigma_{g}^{2}=1$. The algorithm runs on a laptop having an Intel(R) Core(TM) i7-1065G7 CPU 1.30GHz processor with 16GB RAM. The running time per iteration  is around 0.02 seconds. 

The left panel of \Cref{fig:purees_BDRJ} shows the adjacency matrix estimated using the BFDR in \Cref{eqn:BFDR}, where filled boxes represent the selected edges. The right panel displays the corresponding network; nodes are colored according to their group membership. The posterior adjacency matrix is characterized by two large diagonal blocks. The first one represents interactions within the main peak (the three red groups), also revealing short-term interactions between the first group and the one immediately after. Similarly, the second diagonal block represents short-term interactions between the two peaks in the spectrum's tail and their connections.

However,  a distinctive feature of graphical models is the possibility of investigating long-term interactions, and we indeed found some off-diagonal blocks. The most connected group is the red one, $[1018.17 - 1089.48] cm^{-1}$, which is the central part of the main peak. It has four long-term interactions, and, in particular, it is connected to both the final peaks. Another connection is the one between the first peak and the second-to-last peak (in yellow). It is the only long-term connection that does not involve the main peak nor the last one. Finally, note that four nodes in the interval $[1303.41 - 1398.48] cm{-1}$ are completely disconnected, which is probably due to the fact that curves are  flat in that interval, meaning that no particular substance is absorbing in that region. Therefore, it makes sense that it is not correlated with the others. Such a disconnected component is also reported in \citet{codazzi2021gaussian}. 

Our estimated structure of dependencies and the one reported in \citet{codazzi2021gaussian} are similar, even though there are some differences due to the different modeling assumptions. The work of \citet{codazzi2021gaussian} relied on the BDgraph method, which does not look for block structured graphs. Therefore the estimated graph is more fragmented, i.e.,  there are incomplete extra diagonal blocks and some isolated edges. Moreover, the graph is sparser than the one shown in Figure \ref{fig:purees_BDRJ}. This is coherent with the simulation experiments performed in \Cref{section:simulation}, where we empirically showed that BDgraph is more parsimonious in including edges, which leads to a lower number of false positiveness but also to a lower number of discoveries with respect to BDRJ. Moreover, the approximation of the G-Wishart normalizing constants used in BDgraph can also be a cause of increased sparsity, as pointed out by the authors \citep{mohammadi2017ratio}. On the other hand, our method is, by construction, prone to false positiveness. This difference is clearly visible when analyzing the dependencies in the tail of the spectrum, $[1541.1 - 1802.56] cm^{-1}$. The groups are large, with five and seven nodes, respectively. According to BDgraph, some connections among the nodes in such groups are present, but only a few are selected. BDRJ is also able to recognize such interactions. However, it includes a large diagonal block of nodes due to the size of the involved groups.
We notice that also the contrary is possible: BDgraph found some isolated edges that, according to BDRJ, are not strong enough to justify the inclusion of a whole block and therefore are filtered away.

Overall, our approach improves the interpretability of the research findings. We recall that in an absorbance spectrum, each chemical group absorbs light at a specific wavelength; therefore, the main goal is to study the interactions between the peaks of the spectrum, as suggested by the experts. As a consequence, the BDgraph solution is overly detailed since incomplete blocks do not provide information about the relationships between the peaks but only about each portion of the peaks; isolated edges are even more difficult to explain. 
On the other hand, our model takes into account the prior knowledge and balances the
mathematical exactness and the interpretability of the results.

%% file: 6-Discussion.tex
\section{Discussion}
\label{section:discussion}
We introduced a new class of priors, called block graph priors, that allows us to include, in a Gaussian graphical model, prior knowledge available about a partition of the nodes. 
A novel representation of block structured graphs is presented and then exploited to build the prior distribution on the space of block graphs. 

Posterior inference for general non-decomposable graphs is carried out under the G-Wishart prior through a novel sampling strategy. The BDRJ algorithm is a trans-dimensional MCMC sampler on the joint space of graphs and precision matrices that, at each iteration, jointly modifies  an arbitrary number of edges. With a double reversible move, the algorithm avoids the calculation of the normalizing constant $I_G(b,D)$, thus overcoming the main inferential challenge with G-Wishart priors. 
Alternative approaches include \citet{Tan2017} who, however, sequentially update the edges within a randomly selected group of nodes; \citet{willemBlocks} that use a Laplace approximation of the G-Wishart normalizing constant, which is suitable only when $p<20$; and \citet{cremaschi2021seemingly} that rely on a modification of the Birth and Death algorithm of \citet{mohammadi.wit2015}, who apply an exchange algorithm.
Moreover, thanks to  the dimensional reduction of the graph space, the BDRJ algorithm is usually able to get sharper solutions, in particular in terms of the $\text{F}_{1}\text{-score}$ index, relative to unstructured-graph stochastic search approaches.

We recall that our model is grounded on the hypothesis that variables in different groups are fully connected or not connected at all. This modeling hypothesis is assumed to 
encode prior information about the groups of nodes. One side effect of such an assumption is that we can no longer identify the precise structure within each block. 
However, as soon as the dimension of the graph increases, this modeling feature leads to an easier interpretation of the graph estimates and helps to understand  the global behavior of the phenomenon generating the data, rather than looking for all possible, even meaningless, one-to-one relationships among the variables. 
 
A limitation of the current sampling strategy is the lower acceptance rate compared to stochastic search methods based on local moves. The issue can be mitigated by running the algorithm for many iterations at the cost of an additional computational burden.  
Also, a natural concern regards 
the need of tuning the $\sigma^{2}_{g}$ parameter, which defines the perturbation of the elements of the precision matrix that are modified in the construction of the proposed state. This parameter plays a key role in the definition of the chain, and its value has to be fixed a priori. 
A possible solution is to use an adaptive scheme or to generalize the recent 
sequential Monte Carlo method of \citet{willemUnbiased} to the case of global moves, i.e., when an arbitrary set of edges of the graph is changed. 

A further extension can consider other types of graphical models,
such as log-linear models or  non-Gaussian data by using a copula transition and extending them to the framework of block structured graphs.

Finally, in this work, we considered the case of groups of nodes that are contiguous and known a priori. In a more general framework, one could be interested in also learning the partition of the nodes. 
Some works have been proposed in this direction, both using penalized likelihood methods 
\citep{Ambroise2009, tan2015cluster, kumar2020unified} or stochastic block model-based priors for the graph 
\citep{Palla2012,Sun2015}. The common idea in these works is to induce a clustering of the nodes such that variables are more likely to be connected within each group  than to variables in different groups. Conversely, the main focus of the  method proposed in this paper is to investigate the dependencies among different groups.   
In other words, discovering edges among different groups is our primary interest, while these edges are discouraged in the existing literature.

Recently, \citet{willemBlocks} applied a similar reasoning to the $\operatorname{G-Wishart}$ framework. The authors used a stochastic block model as a prior for the graph, where nodes in a block are assumed to form a clique. However, their sampling strategy relies on the Laplace approximation of the G-Wishart normalizing constant  \citep{moghaddam2009accelerating}, and it cannot be applied to graphs with a fixed labeling of the nodes. A modification of the BDRJ scheme may be, instead, a valid alternative for Bayesian learning of random block structures of graphs.

%% file: ackno.tex
\section*{Supplementary Materials}
\begin{description}
\item[Appendix] The appendix provides the details on how to derive the acceptance-rejection probability of the BDRJ algorithm and its schematic description (appendix.pdf)
\item[R-package BGSL:] R-package BGSL containing code implementing the BDRJ algorithm described in the article. The package also contains all datasets used as examples in the article. (BGSL-main.zip)
\end{description}

\section*{Acknowledgments}
We would like to thank the Editor, the Associate Editor and two Referees for the insightful and constructive comments. The author thank Alessandra Guglielmi for valuable suggestions. Moreover, the author thank Matteo Gianella and Laura Codazzi for the useful discussion about the real data analysis. 


\section*{Disclosure Statement}
The authors report there
are no competing interests to declare.

%% file: Supplementary.tex

\newpage
\clearpage
\pagenumbering{arabic}
\renewcommand*{\thepage}{\arabic{page}}
\appendix
\begin{center}
   \Large Appendix to\\ \textbf{Learning Block Structured Graphs in Gaussian
Graphical Models}
\end{center}
\begin{center} \large Alessandro Colombi$^1$, Raffaele Argiento$^2$, Lucia Paci$^3$ and Alessia Pini$^3$\\\medskip
\normalsize
$^1$Department of Economics, Management and Statistics, Università degli studi di Milano-Bicocca\\
$^2$Department of Economics, Università degli studi di Bergamo\\
$^3$Department of Statistical Sciences, Università Cattolica del Sacro Cuore.
\end{center}
\section{Deriving the acceptance-rejection probability}
\label{section:ARMH}

The BDRJ algorithm considers switching between $(\bmK\ats,G\ats,\bmwidetW,G')$ to the alternative $(\bmK',G',\bmW^0,G\ats)$, where $\bmW^0\in\P_{G\ats}$, by performing two Reversible Jump moves: (i) a dimension increasing step from $(\bmK\ats,G\ats)$ to $(\bmK',G')$ according to posterior parameters $b+n$ and $D+U$ and (ii) a dimension decreasing step from $(\bmwidetW, G')$ to $(\bmW^0, G\ats)$ according to prior parameters $b$ and $D$. 

As mentioned in \Cref{section:graph}, the proposed graph $G'$ is obtained from Equation \eqref{eqn:proposal_G} by adding the edge $(l,m)$ to the  multigraph representation of $G^{[s]}$. 
Regarding the precision matrix,  
 the double reversible jump is performed by  leveraging on the change of variables $\bmK \mapsto \bmPhi$, where $\bmPhi$ is an upper triangular matrix such that $\bmK = \bmPhi\T\bmPhi$, see \Cref{section:proposing_K}. 
We set $\bmPhi'_{ij} = \bmPhi_{ij}$ for all $(i,j)\in\nu\left(G^{\ats}\right)$. The free elements are the ones in the set $L$ that are proposed by perturbing the old values independently and with the same variance $\sigma^2_g$, which is a tuning parameter. 
Namely, we draw $\eta_{h}\mytilde{ind}\mbox{N}\left(\bmPhi_{h}^{[s]}, \sigma^{2}_{g}\right)$ and set $\bmPhi'_{h}=\eta_{h}$ for each $h\in L$. 
This defines all the free elements of $\bmPhi'$, while the non-free elements are determined through the completion operation \citep{a.ataykayish.massam2005}. Hence,  $\bmPhi'$ is well defined as well as $\bmK' = (\bmPhi')\T\bmPhi'$. 

The probability to accept the proposed values of $(\bmK',G')$ is equal to 
$\text{min}(1,R^+)$, where
\begin{equation}
    R^+ = \tilda
    \frac{p\left(\bmK',G',\bmW^0,G\ats \mid \mathbf{y}\right)}{p\left(\bmK\ats,G\ats,\bmwidetW,G'\mid \mathbf{y}\right)} \tilda
    \frac{J\left(\bmK'\rightarrow \bmPhi'\right)J\left(\bmW^0 \rightarrow \bmPhi^0\right)}
    {J\left(\bmK\ats\rightarrow \bmPhi\ats\right)J\left(\bmwidetW \rightarrow \bmwidetPhi\right)}
    \tilda
        \frac{q\left(\bmK'\mid\bmK\ats\right)}{q\left(\bmW^0\mid\bmwidetW\right)},
    \label{eqn:apx_BDRJ1}
\end{equation}
where $J(A\rightarrow B)$ denotes the Jacobian of the transformation from A to B. 
As usual with discrete spaces, in Equation \eqref{eqn:apx_BDRJ1}, the Jacobian needed for matching the dimensions of the compared states has been omitted since it reduces to the determinant of the identity matrix.  

First, we recall that the Cholesky decomposition of the precision matrix discussed in \Cref{section:proposing_K} allows us to easily compute the determinant of $\bm{K}$, see \citet{roveratoa.2002},  that is 

\begin{equation}
	\text{det}(\bm{K}) = \prod_{i = 1}^{p}\bmPhi_{ii}^{2}.
	\label{eqn:apx_BDRJ_detK}
\end{equation}
Note that this formulation involves only diagonal values of $\bmPhi$, which are free elements by definition. Hence, Equation  \eqref{eqn:apx_BDRJ_detK} implies that $\text{det}(\bm{K}') = \text{det}\left(\bm{K}^{[s]}\right)$.

The first ratio in Equation  \eqref{eqn:apx_BDRJ1} can be factorized   
as follows:
\begin{nalign}
    \frac{p\left(\bmK',G',\bmW^0,G\ats \mid \mathbf{y}\right)}{p\left(\bmK\ats,G\ats,\bmwidetW,G' \mid \mathbf{y}\right)} 
    \tilda  & =   
    \frac{p\left(\mathbf{y}\mid\bmK',G'\right)}{p\left(\mathbf{y}\mid\bmK\ats,G\ats\right)}
    \tilda
    \frac{p(\bmK'\mid G')}{p\left(\bmK\mid G\ats\right)}
    \tilda
    \frac{q\left(G\ats\mid G'\right)}{q\left(G'\mid G\ats\right)} \\
    & \times
    \frac{p\left(\bmW^0\mid G\ats\right)}{p\left(\bmwidetW\mid G'\right)}
    \tilda 
    \frac{\pi(G')}{\pi\left(G\ats\right)}\tilda \\
    & = 
    \frac{\sqrt{|\bmK'|}}{\sqrt{|\bmK|}} \ 
    \exp\left\{ -\frac{1}{2} \left\langle\bm{K}'-\bm{K}^{[s]}, \ U\right\rangle  \right\} \\
    &  \times
    \frac{I_{G\ats}(b,D)}{I_{G'}(b,D)}
    \exp\left\{ -\frac{1}{2} \left\langle\bm{K}'-\bm{K}^{[s]}, \ D\right\rangle  \right\} \frac{\left\lvert nbd^{\CcB,+}_{M}\left(G^{[s]}_{B}\right) \right\lvert}
    {\left\lvert nbd^{\CcB,-}_{M}\left(G'_{B}\right) \right\lvert}\\
    & \times 
    \frac{I_{G'}(b,D)}{I_{G\ats}(b,D)}
    \frac{1}{\exp\left\{ -\frac{1}{2} \left\langle\widetilde{\bm{W}}-\bm{W}^{0},~D\right\rangle \right\}}
    \frac{\pi(G')}{\pi(G\ats)}
    \\
    &= 
    \frac{\exp\left\{ -\frac{1}{2} \left\langle\bm{K}'-\bm{K}^{[s]},~D+U\right\rangle  \right\}}
    {\exp\left\{ -\frac{1}{2} \left\langle\widetilde{\bm{W}}-\bm{W}^{0},~D\right\rangle \right\}
    } \
    \frac{\left\lvert nbd^{\CcB,+}_{M}\left(G^{[s]}_{B}\right) \right\lvert}
    {\left\lvert nbd^{\CcB,-}_{M}\left(G'_{B}\right) \right\lvert} \
    \frac{\pi(G')}{\pi\left(G^{[s]}\right)},
    \label{eqn:apx_BDRJ_a}
\end{nalign}
where  $\langle A,B\rangle$ denotes the trace of the product between $A$ and $B$. 
Note that the two ratios of $\operatorname{G-Wishart}$ densities allow us to eliminate the presence of their normalizing constants. 
Also, 
note that, thanks to Equation \eqref{eqn:apx_BDRJ_detK}, all the determinants of the matrices in Equation \eqref{eqn:apx_BDRJ_a} canceled out.

For what concerns the change of variable from a precision matrix to its Cholesky decomposition, 
the Jacobian of such a transformation is 

\begin{equation}
    J(\bmK \mapsto \bmPhi) = 2^p\prod_{i=1}^{p}\bmPhi_{ii}^{\nu_i^G},
    \label{eqn:apx_BDRJ_Jacobian}
\end{equation}
where $\nu_{i}^{G}=\lvert\{j: j>i\text{ and }(i,j)\in E \}\lvert$
is the sum of elements in $i$-th row of the adjacency matrix, from position $i+1$ up to the end.
Then, the ratio of the Jacobians appearing in Equation \eqref{eqn:apx_BDRJ1} is readily computed using Equation \eqref{eqn:apx_BDRJ_Jacobian}. That is,
\begin{nalign}
    \frac{J(\bmK'\rightarrow \bmPhi')}{J(\bmK\ats\rightarrow \bmPhi\ats)} 
    \tilda = \tilda 
    \frac{2^p}{2^p} \tilda 
    \frac{\prod_{i=1}^{p}\left(\bmPhi'_{ii}\right)^{\nu_i^{G'} + 1}}{\prod_{i=1}^{p}\left(\bmPhi\ats_{ii}\right)^{\nu_i^G + 1}} 
    \tilda = \tilda 
    \prod_{i\in V(L)}\left(\bmPhi\ats_{ii}\right)^{\nu^{G'}_{i} - \nu^{G^{[s]}}_{i}}.
    \label{eqn:apx_BDRJ_jacobian_ratio}
\end{nalign}
The last equality follows by noticing that  $\nu_i(G) = \nu_i(G')$ for all $i\neq V(L)$ and  those diagonal elements are not modified by construction.
Analogously, one can show that
$$
\frac{J(\bmW^0\rightarrow \bmPhi^0)}{J(\bmwidetW\rightarrow \bmwidetPhi)} 
\tilda = \tilda 
\prod_{i\in V(L)}\left(\bmPhi^0_{ii}\right)^{\nu^{G'}_{i} - \nu^{G^{[s]}}_{i}}.
$$ 
Under the assumption that $G'$ is obtained by adding edge $(l,m)$ to the  multigraph representation of $G^{[s]}$, the exponent in \eqref{eqn:apx_BDRJ_jacobian_ratio} reduces to 
$$
    \nu^{G'}_{i} - \nu^{G^{[s]}}_{i} = \lvert\{j\in B_m : j > i \}\lvert,
    \label{eqn:apx_BDRJ_nui}
$$
which is equal to the number of nodes in group $m$ whose index is greater than $i$.


Finally, the last ratio in Equation \eqref{eqn:apx_BDRJ1} is due to the randomness in the construction of the proposed and the auxiliary matrices. By definition, each term is just the ratio of independent multivariate Gaussian densities, i.e.,
\begin{equation*}
    q\left(\bmK'\mid \bmK\ats\right) =
    \left(\frac{1}{\sqrt{2\pi\sigma^2_g}}\right)^{|L|}
    \exp\left\{
    -\frac{1}{2\sigma^2_g}\sum_{h\in L}\left(\bmPhi'_h - \bmPhi_h\right)^2
    \right\},
    \label{eqn:apx_BDRJ_qK}
\end{equation*}
where, for sake of clarity,  we explicitly wrote the ratio in terms of $\bmPhi'$. 
Similarly, we obtain the quantity $q\left(\bmW^0\mid\bmwidetW\right)\propto \exp\left\{
    -\frac{1}{2\sigma^2_g}\sum_{h\in L}\left(\bmPhi^0_h - \widetilde\bmPhi_h\right)^2
    \right\}$. 

Wrapping everything together, we end up with
\begin{nalign}
    R^{+} &=
    \frac{\exp\left\{ -\frac{1}{2} \left\langle\bm{K}'-\bm{K}^{[s]},~D+U\right\rangle  \right\}}
    {\exp\left\{ -\frac{1}{2} \langle\widetilde{\bm{W}}-\bm{W}^{0},~D\rangle \right\}}~
    \prod_{i\in V(L)}\left(~\frac{\bmPhi^{[s]}_{ii}}{\bmPhi^{0}_{ii}}~\right)^{\nu^{G'}_{i} - \nu^{G^{[s]}}_{i}}\\
    & \times\exp\left\{~\frac{1}{2\sigma^{2}_{g}}~\sum_{h\in L}\left[\left(\bmPhi'_{h} - \bmPhi^{[s]}_{h}\right)^{2} - \left(\bmPhi^{0}_{h} - \bmwidetPhi_{h}\right)^{2}\right] ~\right\} \frac{\pi(G')}{\pi\left(G^{[s]}\right)}.\nonumber
    \label{eqn:apx_BDRJ_final}    
\end{nalign}

\begin{algorithm}
    \SetAlgoLined
    Suppose the chain to be in state $\left(\bm{K}^{[s]},G^{[s]}\right)$, with $\bm{K}^{[s]}=\left(\bmPhi^{[s]}\right)^\top \left(\bmPhi^{[s]} \right)\in P_{G^{[s]}}$ and $G^{[s]} \in \mathcal{B}$.\\ For each iteration:\
    \begin{itemize}
  
    		\item[\textbf{Step 1.}] Updating the graph $G$
    		\begin{itemize}
    		    \item[1.1.] Sample $G_B'$ from $q\left(G_B'\mid G^{[s]}\right)$ given by \eqref{eqn:proposal_G}. Set $G' = \rho^{-1}\left(G^{[s]}\right)$.
    		    Suppose an addition move is selected. 
    		    Call $L$ the set of new edges.
    		    \item[1.2.] Draw $\widetilde{\bm{W}}\mid G'~\sim~\operatorname{G-Wishart}(b,D)$ from an exact sampler \citep{alexlenkoski2013}.
    		    \item[1.3.] For each $h\in L$, draw $\eta_{h} \sim N\left(\bmPhi^{[s]}_{h}, \sigma^{2}_{g}\right)$
    		    \item[1.4.] Set $(\bmPhi')^{\nu\left(G^{[s]}\right)} = \left(\bmPhi^{[s]}\right)^{\nu\left(G^{[s]}\right)}$ and $\bmPhi'_{h} = \eta_{h}~~\forall h \in L$.\\
    		    Derive the remaining elements by completion operation and define $\bm{K}'=(\bmPhi')^{\top}\bmPhi'$.
    		    \item[1.5.] Set $(\bmPhi^{0})^{\nu\left(G^{[s]}\right)} = (\bmwidetPhi)^{\nu\left(G^{[s]}\right)}$. Derive the remaining elements by completion operation and define $\bm{W}^{0}=\left(\bmPhi^{0}\right)^{\top}\bmPhi^{0}$.
    		    \item[1.6.] Compute $\gamma\left(\left(\bm{K}^{[s]},G^{[s]}\right)\rightarrow \left(\bm{K}',G'\right)\right) = \text{min}\{1, R^{+}\}$ where
    		    \begin{nalign}\nonumber
                    R^{+} =~&
                    \frac{\exp\left\{ -\frac{1}{2} \left\langle\bm{K}'-\bm{K}^{[s]},~D+U\right\rangle  \right\}}
                    {\exp\left\{ -\frac{1}{2} \left\langle\widetilde{\bm{W}}-\bm{W}^{0},~D\right\rangle \right\}}~
                    \prod_{i\in V(L)}\left(~\frac{\bmPhi^{[s]}_{ii}}{\bmPhi^{0}_{ii}}~\right)^{\nu^{G'}_{i} - \nu^{G^{[s]}}_{i}}\\
                    &\times \exp\left\{~\frac{1}{2\sigma^{2}_{g}}~\sum_{h\in L}\left[\left(\bmPhi'_{h} - \bmPhi^{[s]}_{h}\right)^{2} - \left(\bmPhi^{0}_{h} - \bmwidetPhi_{h}\right)^{2}\right] ~\right\}\frac{\pi(G')}{\pi\left(G^{[s]}\right)}.
            \label{eqn:acceptance_prob_Block_EA}    
                \end{nalign}
                
    		    \item[1.7.] Draw $c \sim \text{Unif}[0,1]$. if $c < \gamma $ then  set $G^{[s+1]} = G'$.
    		\end{itemize}
    		       	\item[\textbf{Step 2.}] Updating the precision matrix $\mathbf{K}$\\ 
Draw $\bm{K}^{[s+1]} \mid G^{[s+1]}, \mathbf{y} \sim  \operatorname{G-Wishart}(b+n,D+U)$.
    	\end{itemize}
 	\caption{Block Double Reversible Jump}
 	\label{algo:BlockEA}
\end{algorithm}